\documentclass{sigchi}
\usepackage[utf8]{inputenc}
\toappear{
Permission to make digital or hard copies of all or part of this work for personal or classroom use is granted without fee provided that copies are not made or distributed for profit or commercial advantage and that copies bear this notice and the full citation on the first page. Copyrights for components of this work owned by others than ACM must be honored. Abstracting with credit is permitted. To copy otherwise, or republish, to post on servers or to redistribute to lists, requires prior specific permission and/or a fee. Request permissions from permissions@acm.org. \\ 
{\confname{CSCW'16}}, March XX--XX 2016, San Francisco, CA, USA.\\
Copyright is held by the authors. Publication rights licensed to ACM.\\
ACM 978-1-4503-2922-4/15/03...\$15.00.\\ 
}

\usepackage{multirow}
\usepackage{booktabs}
\usepackage{array}
\usepackage{balance}  
\usepackage{graphics} 
\usepackage{times}    
\usepackage{url}      
\usepackage[usenames,dvipsnames,table]{xcolor} 
\usepackage{tikz}
\usetikzlibrary{arrows}

\makeatletter
\def\url@leostyle{%
  \@ifundefined{selectfont}{\def\UrlFont{\sf}}{\def\UrlFont{\small\bf\ttfamily}}}
\makeatother
\newcolumntype{L}{>{\centering\arraybackslash}m{3cm}}

\def\pprw{8.5in}
\def\pprh{11in}

\usepackage[pdftex]{hyperref}
\hypersetup{
pdftitle={Analyzing Organizational Routines in Online Knowledge Collaborations: A Case for Sequence Analysis in CSCW},
pdfauthor={LaTeX},
pdfkeywords={Wikipedia; peer production; online knowledge collaboration; sequence analysis; socio-technical system; organizational practice; routines},
bookmarksnumbered,
pdfstartview={FitH},
colorlinks,
citecolor=black,
filecolor=black,
linkcolor=black,
urlcolor=black,
breaklinks=true,
}

\begin{document}

\title{Analyzing Organizational Routines in Online Knowledge Collaborations: A Case for Sequence Analysis in CSCW}

\numberofauthors{1}
\author{
	\alignauthor Brian C. Keegan,$^{1}$ Shakked Lev,$^{2}$ Ofer Arazy $^{2,3}$\\
     \affaddr{$^{1}$ Harvard Business School, Boston, Massachusetts, USA}\\
     \affaddr{$^{2}$ Department of Information Systems, University of Haifa, Haifa, Israel}\\
     \affaddr{$^{3}$ School of Business, University of Alberta, Edmonton, Alberta, Canada}\\
     \email{bkeegan@acm.org}, \email{levshakked@gmail.com}, \email{oarazy@is.haifa.ac.il}
}

\clubpenalty = 100000000
\widowpenalty = 100000000

\maketitle

\begin{abstract}
Research into socio-technical systems like Wikipedia has overlooked important structural patterns in the coordination of distributed work. This paper argues for a conceptual re-orientation towards sequences as a fundamental unit of analysis for understanding work routines in online knowledge collaboration. We outline a research agenda for researchers in computer-supported cooperative work (CSCW) to understand the relationships, patterns, antecedents, and consequences of sequential behavior using methods already developed in fields like bio-informatics. Using a data set of 37,515 revisions from 16,616 unique editors to 96 Wikipedia articles as a case study, we analyze the prevalence and significance of different sequences of editing patterns. We illustrate the mixed method potential of sequence approaches by interpreting the frequent patterns as general classes of behavioral motifs. We conclude by discussing the methodological opportunities for using sequence analysis for expanding existing approaches to analyzing and theorizing about co-production routines in online knowledge collaboration.
\end{abstract}

\keywords{Wikipedia; peer production; online knowledge collaboration; sequence analysis; socio-technical system; organizational practice; routines}

\category{H.5.3}{Information Interfaces and Presentation}{Group and Organization Interfaces (collaborative computing, computer supported cooperative work)}

\section{Introduction}
Once upon a time, organizational researchers trafficked in platitudes that ``X is an organization that works in practice, but not in theory'' where X might be Linux, Wikipedia, GitHub, or many other ``open'' production systems. Far from being boundary cases with little ``real world'' relevance, the replication of these models' success across domains outside of software and encyclopedias has prompted the re-evalaution of many core economic, psychological, and social frameworks for organizations. In particular, concepts like ``online knowledge collaboration'' and ``commons-based peer production'' define organizational models where distributed members self-organize to create knowledge-based goods in the absence of hierarchies or markets~\cite{Benkler2006wealth,faraj2011knowledge}.

While theories and methods from network science have been brought to bear to understand the relationships that structure online knowledge collaboration~\cite{agarwal2008digitalsocial,sundararajan_information_2013}, these predominately adopt a static view of collaboration. Conversely, studies exploring the dynamics of how these collaborations emerge and change over time often overlook the role of interaction patterns in structuring these systems. Reconciling both of these streams of research, sequences should become a fundamental unit of analysis for capturing how patterns of interaction unfold over time into coherent structures capturing co-production routines. Sequence analysis methods have been widely used in biology and sociology for decades, and the richness and scale of event log data available to HCI, CSCW, and information system researchers naturally lend themselves to such sequence analysis. In addition to adapting these quantitative methods from enumerating and comparing sequence patterns, these sequences likewise invite qualitative inquiry to contextualize their prevalence and interpret their role in structuring socio-technical behavior. Despite its mixed methods lineage and potential, sequence methods have largely been overlooked for understanding questions about organizational routines within online knowledge collaborations. 

Complex organizational practices are composed of more basic behavioral patterns that are in many ways analogous to the basic genetic sequences encoding complex proteins~\cite{gaskin_toward_2014, yoo_digital_2012}. Socio-technical systems like Wikipedia contain abundant event logs encoding sequence data about the series of revisions made to artifacts (\textit{i.e.}, articles) by individuals (\textit{i.e.}, editors). Research into the temporal dynamics of online knowledge collaboration has received sustained attention across scholarship in HCI and information systems. Largely speaking, this research has employed two primary approaches for capturing system dynamics: (1) ``aggregating'' sequence logs to represent a particular process, creating multiple “snapshots” of the process over time~\cite{van2005discovering}, or (2) ``collapsing'' sequence logs to calculate a pair-wise relationship between entities (e.g. affiliation between users), often forming a network of these entities and using social network analysis (SNA) techniques to study the structure of the network~\cite{jurgens2012temporal, keegan_staying_2012, goggins_social_2010}. While our interest is analyzing the sequences of production and coordination activities that make up work routines, this prior work aggregates and collapses event sequences, thus obscuring the sequential ordering and structures in longitudinal records of co-production activities. In this paper we offer a general approach for representing and analyzing event log data from peer production systems that employs methods from sociology and bio-informatics in order to track behavioral motifs and identify the routines of collaborative work. 

In the following section, we review prior work from sociology and organizational theory around organizational routines to understand the dynamics of social change, and contrast this line of scholarship with research that has examined the temporal dynamics of online peer production communities. A research agenda employing sequences within socio-technical systems as a fundamental unit of analysis has the potential to expand theories and methods for both established CSCW topics like recommendation systems and crowdsourcing as well as emerging CSCW topics like online education, citizen science, and crisis informatics.

In order to illustrate our proposed framework, we describe a case study that explores the co-production sequences. Building on theoretical frameworks of generative routines [18], our empirical analysis focuses on the sequences of contributors collaborating to create a Wikipedia article. Using a a representative sample of  93 articles from English Wikipedia, with 37,515 editing activities made by 16,616 distinct contributors, this case explores whether the co-production entails a small cohort of editors who work closely together, or if conversely the production of content within Wikipedia is built upon lengthy editing sessions by the same editor. Our analysis of the behavioral sequences reveal complex and non-random patterns and shed light on the prevalence of behavioral motifs (\textit{i.e.}, generative routines) that organically emerge as contributors make choices about when and how to engage in online knowledge collaboration.  Towards the end of this paper, we outline research questions in these CSCW topics to be explored using sequence analysis methods.


\section{Background and Review}

Sequence analysis shares a rich interdisciplinary literature with psychology, economics, and sociology where it examines patterns within temporally-ordered data, independent variables that influence these patterns, and how these patterns influence dependent variables. Critically, adjacent elements of a sequence are typically not random, but have strong dependencies reflecting topical, causal, or other latent similarities~\cite{abbott_sequence_1995}. Traditional sequence analyses may model the step-by- step transitions between events or analyze the whole sequence of events by measuring distances between them, identifying unique sequences, or matching similar ``families'' of sequences together. Sequence analysis methods have a rich mixed methods history and have been used to model dependencies in participation shifts during conversations~\cite{gibson_taking_2005}, to analyze reciprocated interactions on team performance~\cite{shelly_sequences_1997}, and to suggest new ways to develop social theory~\cite{pentland_narrative_2007}. 

We note that the use of these methods for understanding organizational processes have primarily focused on production and coordination in traditional co-located, synchronous, and small-scale co-located organizations, which are fundamentally different from the context of the current investigation: distributed, asynchronous, and large-scale online knowledge collaborations. Our goal here is to bridge the gap between methodological developments in sequence analysis and their potential to use large-scale event logs around online knowledge collaborations to understand complex social processes. This section reviews prior work on organizational routines and their role in facilitating coordination, and connects it to the research stream investigating temporal dynamics within technology-mediated collaboration, as well as to the theoretical and ethnographic work that aims at understanding patterns of interaction in online knowledge collaboration. 



\subsection{Organizational routines}
Organizational routines are the primary means by which organizations coordinate work~\cite{becker_handbook_2008}. Coordination entails managing dependencies between activities~\cite{malone_interdisciplinary_1994}. Organization design theory suggests that coordination can be achieved through a variety of mechanisms~\cite{gittell_coordination_2004, tushman_information_1978}, but points to organizational routines as the foundation of any work process that involves coordination among multiple actors. Routines refer to repeated patterns of behavior that are bound by rules and customs and that do not change very much between iterations~\cite{feldman_reconceptualizing_2003}. Routines are the product of explicit attempts to design work practices; they are conducive in helping retain organizational history, but they can also be a source of inertia. 

An alternative view emphasizes the emergent and dynamic dimension of routines that are referred to using evolutionary terms like ``adaptation'' or ``mutation~\cite{nelson_winter._2005}.'' Pentland and colleagues~\cite{pentland_grammatical_1995,pentland_organizational_1994} used grammar as an analogy to explain variation in routines: routines consist of rules that allow people to select elements of a repertoire in order to construct a particular sequence of behavior. Feldman~\cite{feldman_organizational_2000} suggested that newly-introduced sequences of action can also lead to changes in rules and procedures. She introduced agency into the notion of routine and viewed routines as improvisational (or ``performative''). This perspective highlights the internal dynamic of routines, where continuous change is the result of participants' reactions to outcomes of previous iterations of the routine. Thus, such routines are not merely repetitive but are generative because they are refined and facilitate knowledge co-production.

In an attempt to reconcile the two perspectives, Feldman and Pentland~\cite{feldman_reconceptualizing_2003,pentland_designing_2008} present an ontology of organizational routines distinguishing between: (a) the ostensive aspect of a routine embodies what we typically think of as the structure; and (b) the performative aspect embodies the specific actions, by specific people, at specific times and places, that bring the routine to life. The structural/ostensive aspect enables people to account for specific performances of a routine; while the dynamic performative/dynamic aspect creates, maintains, and modifies those structural aspects. The relationship between structural and dynamic aspects of routines creates an on-going opportunity for variation, selection, and retention of new practices and patterns of action within routines. Therefore, routines allow organizations to achieve the balance between adaptability and stability, affecting organizations' ability to adapt to changing circumstances.

Studies of organizational routines typically involve ethnographic case studies, offering rich interpretations of specific practices, but a limited capacity to more broadly illuminate the structure and dynamics of routines and patterns. Moreover, the sheer scale of digital trace data within long-term and large-scale socio-technical systems' event logs complicate many traditional quantitative and qualitative research methods~\cite{agarwal2008digitalsocial,lazer2009life}.  Developing a more general framework requires grammatical or lexical approach to explain the diversity of routine elements and sequences across contexts and organizations~\cite{gaskin_toward_2014}. Such a ``grammatical'' approach integrating in-depth field studies with archival digital trace data would permit a mixed methodological approach to triangulating across levels of analysis, cases, and time.

\subsection{Online knowledge collaborations}
Online knowledge collaborations occur in the absence of traditional organizational mechanisms like stable membership, convergence, sustained interaction, or shared goals, but instead rely upon fluid boundaries, norms, participants and interactions~\cite{faraj2011knowledge}. Even so, empirical analyses of online knowledge collaborations have largely adopted a static view of motivations, roles, and routines, ignoring how work unfolds and changes over time in relation to other events~\cite{crowston_freelibre_2008}. The success of peer production systems primarily relying upon distributed, volunteer labor can be understood by examining the artifacts within these collaborations as the focal point where small, self-motivated contributions gradually accumulate and inspire others to contribute in kind~\cite{howison_collaboration_2014}. Collaborators in online knowledge collaborations negotiate tensions between changing and retaining established artifacts within systems by adopting specialized roles, shifting production foci, and employing identifiable action patterns~\cite{kane_emergent_2014}. 

Online knowledge collaborations like Wikipedia and open source development projects rely upon detailed records of contribution and communication patterns. Many socio-technical systems archive records and other meta-data about changes in the state of the system into event logs. These data are valuable for contributors to trace changes across versions of documents~\cite{forte_decentralization_2009}, to evaluate others' contributions~\cite{tsay_influence_2014}, and to build additional tools to support collaboration~\cite{ducheneaut_2005_socialization}. Users' contributions drive the development of artifacts within the system, but these contributions also re-shape the governance mechanisms among developers, portending new roles and capacities within the online community~\cite{de_souza_seeking_2005}. This agency versus structure tension manifests as a changing emphasis between experimenting with new ideas and processes on the one hand~\cite{nakakoji_evolution_2002} and supporting stable organizational routines and social roles on the other~\cite{fisher2004everyday, gleave_conceptual_2009}. Event log data capture these complex interactions, by encoding both the agency of contributors to act across artifacts and respond to others' contributions as well as recording events that structure the community such as the development of rules.\\

\subsection{Sequence analysis}
Following calls to analyze the micro-level processes of peer production systems~\cite{lakhani_how_2003} and despite the reliance on event logs to understand how work processes and roles are structured over time in prior research~\cite{gleave_conceptual_2009}, these analyses have largely ignored the role of detailed sequences in log data and overlooked the temporal and artifact-related dependencies. Prior research investigating co-production dynamics as either aggregated sequence log to represent a particular process, creating multiple ``snapshots'' of the process over time, or ``collapsed'' sequence logs to calculate a pair-wise relationship between entities, forming a network of these entities and using social network analysis techniques to study the structure of the network~\cite{jurgens2012temporal, keegan_staying_2012, goggins_social_2010}. Less relevant to our investigation are prior works that employed information visualization techniques --– rather than a formal knowledge representation --– to depict the dynamics of collaborative work~\cite{dunne_graphtrail_2012,smith_visualization_2001,viegas_studying_2004}.

Few studies have gone further to provide a more detailed account of temporal dynamics within peer-production through the combination of the two approaches highlighted above and by generating a series of temporal snapshots of network structure~\cite{goggins_connecting_2014}. While these prior works capture some aspects of the temporal dynamics underlying peer production, they aggregate and collapse event sequences, rather than analyze patterns within sequences of co-production routines. Another relevant stream of research has analyzed patterns within event sequences, but was restricted to a single phase transitions, for example in studying activity~\cite{yamauchi_collaboration_2000}, contributor~\cite{olson_paying_2010} or role~\cite{arazy_functional_2015} transitions.\footnote{The restriction to a single phase transition is often associated with the analytic approaches employed (e.g. Markov models).} Despite the various approaches for studying temporal dynamics in online production communities~\cite{fisher2004everyday,goggins_connecting_2014,jahnke_dynamics_2010}, research in the area --– in particular, peer production --– has not yet developed approaches and methods for identifying patterns of co-production activities from event log sequences~\cite{gaskin_toward_2014}. Few recent studies have attempted to analyze multi-stage transitions from event log data~\cite{geiger2013sessions} and have explored the application of methods borrowed from bio-informatics~\cite{arazy_sustainability_2010,gaskin_toward_2014,yoo_digital_2012}. Nonetheless, these studies have employed simple knowledge representation schemes, often with an event of a single type.

Our goal in this paper is to extend existing approaches to sequence analysis by proposing a general framework for analyzing multi-stage and multi-type co-production routines, by employing event log sequences and relying on methods common in bio-informatics. Because event logs within online knowledge collaborations encode meta-data about the who, what, when, where, and how of socio-technical activities, an analysis of sequence logs could reveal the routines within artifact coordination and co-production. As we detail in the subsequent section, such event logs could allow researchers to enrich the representation of behavioral changes and organizational structure at multiple levels of analysis while using a mixture of quantitative and qualitative methods. Much as computational methods from natural language processing or social network analysis enable researchers to extract new features from a corpus of unstructured data, we imagine sequence analysis becoming a similar methodological toolkit for researchers to interrogate data. In the remainder of the paper, we use a case study to explore the potential of quantitatively mining and qualitatively interpreting sequence data derived from socio-technical event logs. Generalizing from this case study, we outline a proposed research agenda for employing large-scale event log data to understand the relationships, patterns, predictors, and consequences of sequence data across various organizational phenomena. Finally, we discuss the implications this research agenda has for extending sequence methods to empirically study and theorize routines in technology-mediated collaborations.

\section{Sequence analysis methodology framework}
\begin{table}[tb]
\centering
  \begin{tabular}{cccc}
  \toprule[0.125em]
    \textbf{Activity} & \textbf{Artifact} & \textbf{Performer} & \textbf{Order} \\ 
    \textit{Edit type} & \textit{Article} & \textit{Editor} & \textit{Time} \\
    \cmidrule(lr){1-4}
    Rephrase Text (RT) & X & $U_1$ & 1:01 \\ 
    Change Markup (CM) & X & $U_2$ & 2:02 \\ 
    Add Content (AC) & X & $U_3$ & 3:03 \\ 
    Fix Typos (FT) & X & $U_1$ & 4:04 \\ 
    Add Vandalism (AV) & X & $U_4$ & 5:05 \\ 
    Delete Vandalism (DV) & X & $U_1$ & 6:06 \\ 
  \bottomrule[0.125em]
  \end{tabular}
\caption{Example of an event log.}
\label{table:event_log}
\vspace{-1.0em}
\end{table}

\begin{figure}[tb]
\centering
  \begin{tabular}{c}

    \begin{tikzpicture}
    [user/.style={circle,fill=Black!60,thick,text=white,inner sep=0pt,minimum size=.75cm},
    ->,-stealth,line width = 2.5, black,
    font=\boldmath]
    \node [user] (a1) at (0,0) {$RT$};
    \node [user] (b1) at (1.25,0) {$CM$};
    \node [user] (c1) at (2.5,0) {$AC$};
    \node [user] (a2) at (3.75,0) {$FT$};
    \node [user] (d1) at (5,0) {$AV$};
    \node [user] (a3) at (6.25,0) {$DV$};
    \path
    (a1) edge (b1)
    (b1) edge (c1)
    (c1) edge (a2)
    (a2) edge (d1)
    (d1) edge (a3);
    \end{tikzpicture} \\
  \\
    \begin{tikzpicture}
    [user/.style={circle,fill=Black!60,thick,text=white,inner sep=0pt,minimum size=.75cm},
    ->,-stealth,line width = 2.5, black,
    font=\boldmath]
    \node [user] (a1) at (0,0) {$X$};
    \node [user] (b1) at (1.25,0) {$X$};
    \node [user] (c1) at (2.5,0) {$X$};
    \node [user] (a2) at (3.75,0) {$X$};
    \node [user] (d1) at (5,0) {$X$};
    \node [user] (a3) at (6.25,0) {$X$};
    \path
    (a1) edge (b1)
    (b1) edge (c1)
    (c1) edge (a2)
    (a2) edge (d1)
    (d1) edge (a3);
    \end{tikzpicture} \\
  \\
    \begin{tikzpicture}
    [user/.style={circle,fill=Black!60,thick,text=white,inner sep=0pt,minimum size=.75cm},
    ->,-stealth,line width = 2.5, black,
    font=\boldmath]
    \node [user] (a1) at (0,0) {$U_1$};
    \node [user] (b1) at (1.25,0) {$U_2$};
    \node [user] (c1) at (2.5,0) {$U_3$};
    \node [user] (a2) at (3.75,0) {$U_1$};
    \node [user] (d1) at (5,0) {$U_4$};
    \node [user] (a3) at (6.25,0) {$U_1$};
    \path
    (a1) edge (b1)
    (b1) edge (c1)
    (c1) edge (a2)
    (a2) edge (d1)
    (d1) edge (a3);
    \end{tikzpicture} \\

    \begin{tikzpicture}
    [user/.style={circle,fill=white,thick,text=black,inner sep=0pt,minimum size=.75cm},
    ->,-stealth,line width = 2.5, black,
    font=\boldmath]
    \node [user] (a1) at (0,0) {1:01};
    \node [user] (b1) at (1.25,0) {2:02};
    \node [user] (c1) at (2.5,0) {3:03};
    \node [user] (a2) at (3.75,0) {4:04};
    \node [user] (d1) at (5,0) {5:05};
    \node [user] (a3) at (6.25,0) {6:06};
    \end{tikzpicture}
  \end{tabular}
  \caption{%
  The event log in Table \ref{table:event_log} can be represented as a sequence of activities (top), artifacts (middle), or performers (bottom).}
  \label{fig:three_sequences}
  \vspace{-1.0em}
\end{figure}
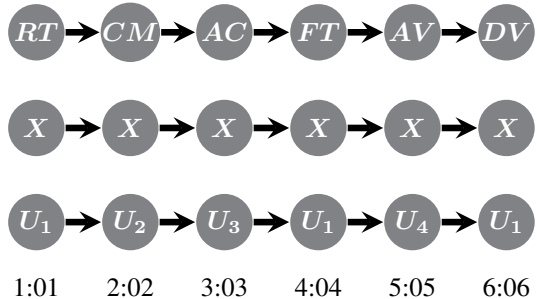 

The availability of detailed sequential event log data in large-scale socio-technical systems provides substantial opportunities to extend interdisciplinary empirical research around routines in new organizational forms. Understanding the distribution, evolution, and generative role of activity sequences within these systems can improve theories about online collective action, distributed work, and shared governance. Socio-technical systems like Wikipedia record the complete history of every change made to an article since the first edit as well as every revision made by any user. Analogous event logs likewise exist for open source repositories containing histories of the commits from every user to a project, and social networking sites archiving the history of actions and posts users have made. These event logs contain distinct records encoding a variety of meta-data related to production (\textit{e.g.}, developing an encyclopedia article), coordination (\textit{e.g.}, discussion around who should do what), user interactions (\textit{e.g.}, comments by one user on another user’s personal page), and governance (\textit{e.g.}, policy enforcement). 

Given the diverse affordances of socio-technical systems, we will bracket our proposed framework by focusing on the modality of users collaborating to create artifacts (\textit{e.g.}, Wikipedia), although alternatives such as users managing relationships to facilitate interpersonal communication (\textit{e.g.}, Twitter) certainly exist. Following on prior work from ``process mining'' that extracts information from event logs to understand how organizations manage business processes~\cite{van2003workflow}, we employ four event log features to construct sequences that capture organizational routines~\cite{van2005discovering}:
\begin{description}
\item[Activity] - A system action such as a revision to an article.\vspace{-.5em}
\item[Artifact] - An entity which is acted upon such as an article.\vspace{-.5em}
\item[Performer] - An entity executing the activity such as a user.\vspace{-.5em}
\item[Order] - An index defining a sequence such as a timestamp.
\end{description} 

A simple example of an event log is given in Table~\ref{table:event_log} for contributions (``commits'') from a variety of users $\{U_1,U_2,U_3,U_4\}$ to a single artifact $X$. As a generalizable framework, we construct activities, artifacts, performers, and order broadly. Activities are not merely edits or commits, but the granting or revocation of permissions, creation of new relationships, or instances of communicating between users. Artifacts are not merely information-based products, but may also refer to coordination elements of various types like discussion or email threads or governance repositories logging the history of permission, promotion, or enforcement actions. Performers are not only single human users, but potentially sub-users within an account (\textit{e.g.}, characters in an online game) or automated users (\textit{e.g.}, anti-vandalism bots on Wikipedia). Alternative frameworks emphasizing interpersonal communication and relationships communications might employ ego-to-alter elements as opposed to the user-to-artifact constructs as well. 

\subsection{Step 1: Identification of sequences}

This event log of ``users taking actions on an object'' encodes at least three kinds of sequences from Table~\ref{table:event_log}: (1) the sequence of performers (\textit{e.g.}, users) who acted ($U_1$--$U_2$--$U_3$--$U_1$--$U_4$--$U_1$), (2) the sequence of activities (\textit{e.g.}, commits) that were taken ($RT$--$CM$--$AC$--$FT$--$AV$--$DV$), and (3) the sequence of artifacts (\textit{e.g.}, articles from the set ${X}$) that were acted upon ($X$--$X$--$X$--$X$--$X$--$X$). Each of these three sequence types are trivial when looking at the event log of a single performer, the event log of a single activity, and the event log of a single artifact, respectively. For example, Table~\ref{fig:three_sequences} represents a simple case of an event log related to a single artifact, where there is meaningful variation in the sequence of activities (edit types) and of performers (users), but no variation in artifacts (articles). Crucially, these three sequences types have relationships with each other and may also re-occur at other times within the same artifact’s event log as well in other artifacts’ event logs. Figure~\ref{fig:three_sequences} describes the three types of sequences that could be derived form the example in Table~\ref{table:event_log}.

\subsection{Step 2: Schematization of sequences}
\begin{figure}[tb]
\centering
  \begin{tabular}{c}

    \begin{tikzpicture}
    [user/.style={circle,fill=Black!60,thick,text=white,inner sep=0pt,minimum size=.75cm},
    ->,-stealth,line width = 2.5, black,
    font=\boldmath]
    \node [user] (a1) at (0,0) {$U_1$};
    \node [user] (b1) at (1.25,0) {$U_2$};
    \node [user] (c1) at (2.5,0) {$U_3$};
    \node [user] (a2) at (3.75,0) {$U_1$};
    \node [user] (d1) at (5,0) {$U_4$};
    \node [user] (a3) at (6.25,0) {$U_1$};
    \path
    (a1) edge (b1)
    (b1) edge (c1)
    (c1) edge (a2)
    (a2) edge (d1)
    (d1) edge (a3);
    \end{tikzpicture} \\
    \\
    \begin{tikzpicture}
    [user/.style={circle,fill=Maroon!60,thick,text=black,inner sep=0pt,minimum size=.75cm},
    ->,-stealth,line width = 2.5, black,
    font=\boldmath]
    \node [user,fill=orange!50] (a1) at (0,0) {$E$};
    \node [user,fill=orange!50] (b1) at (1.25,0) {$E$};
    \node [user,fill=orange!50] (c1) at (2.5,0) {$E$};
    \node [user,fill=green!50] (a2) at (3.75,0) {$C$};
    \node [user,fill=orange!50] (d1) at (5,0) {$E$};
    \node [user,fill=blue!50] (a3) at (6.25,0) {$B$};
    \path
    (a1) edge (b1)
    (b1) edge (c1)
    (c1) edge (a2)
    (a2) edge (d1)
    (d1) edge (a3);
    \end{tikzpicture} \\

    \begin{tikzpicture}
    [user/.style={circle,fill=white,thick,text=black,inner sep=0pt,minimum size=.75cm},
    ->,-stealth,line width = 2.5, black,
    font=\boldmath]
    \node [user] (a1) at (0,0) {1:01};
    \node [user] (b1) at (1.25,0) {2:02};
    \node [user] (c1) at (2.5,0) {3:03};
    \node [user] (a2) at (3.75,0) {4:04};
    \node [user] (d1) at (5,0) {5:05};
    \node [user] (a3) at (6.25,0) {6:06};
    \end{tikzpicture}
  \end{tabular}
  \caption{%
  The sequence from the example event log in Figure \ref{fig:three_sequences} (top) can be schematized using the rules from Table \ref{table:coding_system} (bottom).}
  \label{fig:translating_to_greek}
  \vspace{-1.0em}
\end{figure}
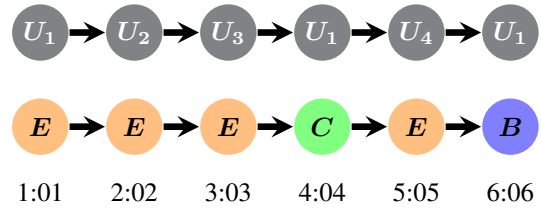

Once the fundamental elements representing online knowledge production have been defined, there is a need to formulate a knowledge representation scheme that would capture the particular sequences of elements that are most relevant for specific research question at hand. For example, questions pertaining to personal relationships between participants and their social networks require that the representation capture participants' identities. In contrast, research questions around the temporal evolution to the community's organizational structure call for representing participants' roles and the timing of role-transition events (but not their identities). We define our knowledge representation scheme in Table~\ref{table:coding_system} with an example provided in Figure~\ref{fig:translating_to_greek}.

\subsection{Step 3: Analysis of sequences}
The third step in our proposed framework calls for the analysis of the extracted sequences. Quantitative methods developed independently in bioinformatics, natural language processing, and sociology have independently developed and extended methods for analyzing similarities and variation patterns of sequence data that could be employed to analyze sequential behavior across  organizations~\cite{gaskin_toward_2014,yoo_digital_2012}. Methods in this vein focus on three classes of analysis:

\begin{description}
\item[Pattern mining.] Using enumerative and deterministic optimization methods to discover the types and frequencies of different sequences within a data set~\cite{dempster_maximum_1977}. The number of sequences and tests for their likelihood against some baseline expectation can reveal important patterns~\cite{han_frequent_2007}. \vspace{-.5em}
\item[Sequence similarity.] A sequence’s similarity to other sequences can be evaluated based on notions of proximity, where more deviations in the elements of one sequence compared to the other implies a larger ``distance''~\cite{dijkstra_measuring_1995, xing_brief_2010}. using techniques such as dynamic time warping  These might be represented as a ``phylogenetic tree'' clustering similar sequences together or ``dynamic time warping'' to represent continuous time series as a sequence of discrete tokens, like letters~\cite{berndt_using_1994,lin_experiencing_2007}.\vspace{-.5em}
\item[Probabilistic analysis.] A sequence can also be the realization of an underlying process that might generate alternative sequences as well~\cite{lafferty_conditional_2001}. Dynamic programming methods and Hidden Markov Models (despite their limitations) analyze sequences to understand the probability of the observed patterns as a result of transitions between finite states~\cite{salamin_introduction_2011}.
\end{description}
Each of these methods reflect different modeling assumptions. While developed for domains  other than online peer production, these tools can generally be applied to sequence data of any kind with substantial potential for generating novel metaphors for understanding complex behavior~\cite{gaskin_toward_2014}.

\subsection{Step 4: Interpretation of sequences}
The final step of our sequence analysis methodology framework is to employ qualitative methods to interpret and triangulate the observed patterns. Such qualitative analysis may rely on secondary and archival data from the online community, interviews with community members~\cite{bryant_becoming_2005}, or a field study~\cite{jahnke_dynamics_2010}. Employing qualitative methods allow for enriching the results obtained through quantitative sequences with a deep understanding of the contextual factors contributing to the observed patterns. Such a mixed method approach can yield richer insights regarding the motivations and contexts for co-production routines. 

\section{Case Study}
\begin{table*}[tb]
  \centering
  \begin{tabular}{c|p{5.5cm}|c|c|c|c|c|c}
    \toprule[1pt]
    \textbf{Label} & \textbf{Description} & \textbf{Count} & \textbf{\%} & \textbf{Size} & \textbf{Duration} & \textbf{Last session} & \textbf{Last same-session} \\
    \toprule[1pt]
	\cellcolor{violet!50} $A$ & The contributor also made the previous editing session in this article & 2,228 & 8\% & 705 & 3.15 & 2.00 & 2.00 \\ \hline
	\cellcolor{blue!50} $B$ & The contributor was active in the editing session preceding the last session & 1,835 & 6\% & 1447 & 0.89 & 1.71 & 4.8 \\ \hline
	\cellcolor{green!50} $C$ & The contributor was active in this article 3-5 editing sessions prior	& 1,350 & 5\% & 623 & 0.94 & 4.61 &26.2 \\ \hline
	\cellcolor{yellow!50} $D$ & The contributor was active in this article $\geq$6 editing session prior & 3,944 & 14\% & 812 & 0.75 & 4.64 & 192.4 \\ \hline
	\cellcolor{orange!50} $E$ & The first editing session of the contributor to this article & 19,562 & 68\%  & 553 & 0.75 & 11.66 & n/a \\ \hline
	\cellcolor{red!50} $F$ & The first editing session of the contributor in Wikipedia & 4 & 0\% & 429 & 1.25 & 6.25 & n/a \\
    \bottomrule[1pt]
  \end{tabular}
  \caption{Coding system. Average session duration is measured in hours, average time since last session and average time since same user last session is measured in days.}
  \label{table:coding_system}
\end{table*}
The focus of this empirical investigation is the English Wikipedia and its community of editors. In contrast to the organization of work within traditional organizations, in the creation of a Wikipedia article each contributor is free to generatively enact a role in the moment~\cite{faraj2011knowledge}. Thus, contributors choose not only when and if to participate the co-production process, they are also free to determine the exact nature of their activity (with very few restrictions, \textit{e.g.} a non-registered member is not allowed to create a new Wikipedia article). The goal of this case study is to explore a particular aspect in co-production routines: the prevalence of different sequences of editing patterns, using the contributor’s ID as the focal point. Such an analysis has the potential to yield insights regarding the number of contributors participating in article co-creation and the sequences of their product-centric interaction (\textit{e.g.}, discovering ``ping-pongs'' between a few contributors working collaboratively to improve the article).

\subsection{Sample and Data Extraction}

Our sample included 93 articles that provide a representation of Wikipedia's topical categories and this sample has been employed in earlier studies~\cite{arazy_information_2011,arazy_measurability_2011}.  The set of articles was selected based on randomization and a stratified sampling of Wikipedia's topics, congruent with Wikipedia's top-level classification~\cite{kittur_he_2007} (categories: culture, art, and religion; math, science, and technology; geography and places; people and self; society; and history and events). From this original set of 96 Wikipedia articles, 3 were discontinued and their edit history is no longer available, leaving us with 93 articles. We tracked all activities to this sample of articles, from each article's inception until December 31, 2012. Altogether, this data set includes 37,515 editing activities made by 16,616 distinct contributors. 

Our investigation is primarily based on a large-scale quantitative analysis of Wikipedia system activity logs, which is complemented with a qualitative analysis. While the quantitative analysis is essential for identifying those patterns that recur frequently and for drawing statistical conclusions, the qualitative analysis allows for a deeper understanding as to why and how co-production unfolds. For the quantitative analyses, we employed data harvested from Wikipedia logs. We queried the Wikipedia API  recording the time and contributor of each edit made to articles in our sample. In addition, we tracked the activity of all contributors who have edited articles in our sample, recording key events in contributors’ Wikipedia career (\textit{e.g.}, from the date of first edit to Wikipedia). For each editing activity in each of the articles in our sample, we recorded the contributor’s ID and the timing of activities. Each article was then represented as a vector of temporally-adjacent sequential editing activities. 

In addition, qualitative analysis was used in order to gain a deeper understanding of a few recurring patterns as examples. For this analysis, we employed a combination of manual and automated methods, relying on both Wikipedia (\textit{e.g.}, using the version compare tool to manually determine the nature of the particular activity) and external resources (\textit{e.g.}, Web activity related to the topic of a particular Wikipedia article, as a means of identifying real-world events that may have triggered Wikipedia editing activity).

\subsection{Data representation}
In order to analyze sequences of activities, it was essential that we devise a consistent way for representing: (a) the fundamental unit of activity; and (b) a sequence of activities. We chose an editing session as the basic unit of analysis, collapsing sequential edits made by the same contributor within a very short period of time~\cite{geiger2013sessions}. Often, contributors would save their edit-in-process, where each saved version is recorded in Wikipedia log as a revision of the article. 

We sought a more meaningful unit of analysis, and collapsed sequential edits within the 10 minutes into s  single edit session. The 10-minute threshold was determined based on a statistical analysis of time-between-edits . The time stamp of the collapsed editing session was defined as the duration between the timing for the first and last edit activities included in that session. Consider the example of a series of edits made by user $U_1$:  $U_1$(10:10)--$U_1$(10:12)--$U_1$(10:21); this edit activity sequence will be collapsed onto the editing session: $U_1$([10:10,10:21]). In investigating sequences in recurring patterns, we process all articles in the data set together. After collapsing all edit activity vectors, 28,923 vectors of editing sessions remained. 

In order to represent sequences of activities (\textit{i.e.}, motifs), it was essential that we devise a novel representation of contributor sequence motifs. Given our focus on the sequences of contributors co-creating the articles, we needed to design a generic motif representation that could represent similar sequences by different contributors. For example, consider the set of four contributors $\{U_1,U_2,U_3,U_4\}$: we wanted to create a single motif to capture the following similar sequences:
\[U_1-U_1-U_2-U_2-U_3\]\vspace{-1.0em}
\[U_2-U_2-U_4-U_4-U_1\]\vspace{-1.0em}
\[U_3-U_3-U_1-U_1-U_2\]
These all represent 2 consecutive editing sessions by the same contributor, followed by two consecutive sessions by a different contributor, then followed by an editing session by a third contributor. 

The questions our representation tries to address are: (1) Who is the person making the most recent edit? (2) Is it the same person who has edited the prior edit? (3) An earlier edit to the same article? (4) Or does this edit represent his first activity in the particular article? While prior works have proposed various approaches for representing co-production motifs in Wikipedia, these motif representations were not appropriate for the more general purpose. Our proposed motif representation traverses the editing session vectors for each article in our sample, starting from the earliest session, and for each editing session records the most recent time that same contributor was active, as seen in Table~\ref{table:coding_system}.

\subsection{Distribution of sequences}
We begin by presenting descriptive statistics for motif labels (see Table~\ref{table:coding_system}). We notice that the vast majority of sessions (68\%) represent a contributor’s first-time activity in the particular article (label $E$), followed by labels representing a non-recent activity by that contributor ($D$; 14\%). The analysis of size of revision shows that revisions by the contributor making the edit session before last ($B$) is the largest (roughly 1500 characters; possibly indicating edit wars, see details below), while those making the first edit to Wikipedia ($F$) tend to make smaller changes. Consecutive edit sessions by the same contributor ($A$) are of a longer duration, indicating that they are often composed of several edits. An analysis of the time between editing sessions reveals that contributors new to the article ($E$) tend to edit after long dormant periods (on average, almost 12 days), contributor making the edit session before last ($B$) often quickly re-act to the previous session (on average, under 2 days; representing ping-pongs between contributors). Finally, an analysis of the times between same-contributor sessions shows that inactive contributors ($D$) often stay away from editing an article for prolonged periods (on average, over six months). 

Table \ref{table:sequence_distributions} below presents the most frequent motifs in lengths of 2-4 labels.\footnote{We analyzed motifs of up to 10-label length, and in essence the patterns observed for these longer motifs were quite similar to the shorter motifs presented in the tables. Because of space limitations, we decided to exclude these longer motifs from this paper.} As can be seen from these results, the vast majority of motifs represent sequences of activities by editors new to the article (label $E$), sometimes interrupted by an editing session by a contributor that was active a while ago ($D$). This is not surprising, given that 68\% of single-label motifs represent new-comers ($E$) and sessions by inactive contributors ($D$) are second most frequent. 

A more telling statistics calculates those motifs that appear substantially more (or less) times than what is expected by chance alone. Assuming that the appearance of a label is independent of the appearance of the previous labels, this is calculated independently for each motif length. That is, the calculation of by-chance occurrence assumes there is no dependency between sequential labels, such that the expected frequency of every motif is simply the multiplication of the frequencies of all its labels. For example, the expected frequency of the motif $B$--$C$--$A$ is 6\% x 5\% x 8\% = 0.024\%. Our analyses identified those motifs that appear significantly above (or below) chance, using a two-tailed $Z$-test. We focus our analyses on the patterns that reoccur frequently in our sample: 2-motif sequences reoccurring at least 200 times; 3-motif sequences with more than 100 occurrences; and 4-motif sequences with more than 50 occurrences. When calculating statistical significance levels for the Z-test, we apply the Bonferroni correction: a statistical method for counteracting the problem of multiple comparisons (family-wise error rate)~\cite{benjamini_controlling_1995}. The Bonferroni correction is very conservative, and comes at the cost of increasing the probability of producing false negatives, and consequently reducing statistical power. Thus, statistical significance after applying this correction suggests that the null hypothesis could be rejected with very high confidence.  

Table~\ref{table:frequent_sequences_tests} below presents the most significant (highest absolute $Z$ scores) for motifs of 2-4 labels (the frequency of all motifs in this table is significantly above/below chance at $p < 0.001$; with Bonferroni correction). As Table~\ref{table:frequent_sequences_tests} reveals, those motifs that appear significantly above/below chance are quite different from the most frequent motifs described in Table \ref{table:sequence_distributions}. We note that while few of the significant motifs are very common (namely sequences of $E$), the other statistically significant motifs are less common. For example, same contributor motifs (label $A$) appear significantly more than what is expected by chance alone: the motif $A$--$A$--$A$ is expected to occur in 0.05\% of 3-label motifs, while it appears 1.69\% of the times; $Z$-score = 131 ($p < 0.001$). 
\newline

\begin{table}[tb]
  \centering
  \begin{tabular}{c|c|c|c|r|r}
    \toprule[1pt]
    \multicolumn{4}{c|}{\textbf{Label}} & \textbf{Count} & \textbf{Fraction} \\
    1st & 2nd & 3rd & 4th & & \\
    \toprule[1pt]
	\cellcolor{orange!50} $E$ & \cellcolor{orange!50} $E$ & & & 14,003 & 48.6\% \\ \hline
    \cellcolor{orange!50} $E$ & \cellcolor{yellow!50} $D$ & & & 2,789 & 9.7\% \\ \hline
    \cellcolor{yellow!50} $D$ & \cellcolor{orange!50} $E$ & & & 2,686 & 9.3\% \\ \hline
	\cellcolor{orange!50} $E$ & \cellcolor{blue!50} $B$ & & & 1,014 & 3.5\% \\ \hline
    \cellcolor{blue!50} $B$ & \cellcolor{orange!50} $E$ & & & 1,003 & 3.5\% \\ \hline
    \cellcolor{orange!50} $E$ & \cellcolor{orange!50} $E$ & \cellcolor{orange!50} $E$ & & 10,469 & 36.4\% \\ \hline
    \cellcolor{orange!50} $E$ & \cellcolor{yellow!50} $D$ & \cellcolor{orange!50} $E$ & & 2,024 & 7.0\% \\ \hline
    \cellcolor{orange!50} $E$ & \cellcolor{orange!50} $E$ & \cellcolor{yellow!50} $D$ & & 1,873 & 6.1\% \\ \hline
    \cellcolor{yellow!50} $D$ & \cellcolor{orange!50} $E$ & \cellcolor{orange!50} $E$ & & 1,742 & 6.1\% \\ \hline
    \cellcolor{orange!50} $E$ & \cellcolor{orange!50} $E$ & \cellcolor{orange!50} $E$ & \cellcolor{orange!50} $E$ & 7,973 & 27.8\% \\ \hline
    \cellcolor{orange!50} $E$ & \cellcolor{orange!50} $E$ & \cellcolor{orange!50} $E$ & \cellcolor{yellow!50} $D$ & 1,397 & 4.9\% \\ \hline
    \cellcolor{orange!50} $E$ & \cellcolor{orange!50} $E$ & \cellcolor{yellow!50} $D$ & \cellcolor{orange!50} $E$ & 1,383 & 4.8\% \\ \hline
    \cellcolor{orange!50} $E$ & \cellcolor{yellow!50} $D$ & \cellcolor{orange!50} $E$ & \cellcolor{orange!50} $E$ & 1,348 & 4.7\% \\ \hline
    \cellcolor{yellow!50} $D$ & \cellcolor{orange!50} $E$ & \cellcolor{orange!50} $E$ & \cellcolor{orange!50} $E$ & 1,227 & 4.3\% \\
    \bottomrule[1pt]
  \end{tabular}
  \caption{Most frequent sub-sequences.}
  \label{table:sequence_distributions}
\end{table}

\begin{table}[tb]
  \centering
  \begin{tabular}{c|c|c|c|r|r|r}
    \toprule[1pt]
    \multicolumn{4}{c|}{\textbf{Label}} & \textbf{Expected} & \textbf{Observed} & \textbf{Z-Score} \\
    1 & 2 & 3 & 4 & & \\
    \toprule[1pt]
	\cellcolor{violet!50} $A$ & \cellcolor{violet!50} $A$ & & & 0.59\% & 2.91\% & 51.2\\ \hline
    \cellcolor{blue!50} $B$ & \cellcolor{blue!50} $B$ & & & 0.40\% & 1.36\% & 25.7\\ \hline
    \cellcolor{orange!50} $E$ & \cellcolor{violet!50} $A$ & & & 5.21\% & 2.83\% & -18.2\\ \hline
	\cellcolor{violet!50} $A$ & \cellcolor{orange!50} $E$ & & & 5.21\% & 3.48\% & -13.2\\ \hline
	\cellcolor{orange!50} $E$ & \cellcolor{orange!50} $E$ & & & 45.74\% & 48.57\% & 9.6\\ \hline
	\cellcolor{violet!50} $A$ & \cellcolor{violet!50} $A$ & \cellcolor{violet!50} $A$ &  & 0.05\% & 1.69\% & 130.8\\ \hline
    \cellcolor{blue!50} $B$ & \cellcolor{blue!50} $B$ & \cellcolor{blue!50} $B$ & & 0.03\% & 0.57\% & 58.2\\ \hline
    \cellcolor{orange!50} $E$ & \cellcolor{orange!50} $E$ & \cellcolor{orange!50} $E$ & & 30.94\% & 36.42\% & 20.1\\ \hline
    \cellcolor{green!50} $C$ & \cellcolor{blue!50} $B$ & \cellcolor{blue!50} $B$ & & 0.02\% & 0.18\% & 19.6\\ \hline   
    \cellcolor{blue!50} $B$ & \cellcolor{violet!50} $A$ & \cellcolor{violet!50} $A$ & & 0.04\% & 0.25\% &18.6\\ \hline
    \cellcolor{violet!50} $A$ & \cellcolor{violet!50} $A$ & \cellcolor{violet!50} $A$ & \cellcolor{violet!50} $A$ & 0.00\% & 1.13\% & 323.5\\ \hline
    \cellcolor{blue!50} $B$ & \cellcolor{blue!50} $B$ & \cellcolor{blue!50} $B$ & \cellcolor{blue!50} $B$ & 0.00\% & 0.29\% & 121.9\\ \hline
    \cellcolor{blue!50} $B$ & \cellcolor{violet!50} $A$ & \cellcolor{violet!50} $A$ & \cellcolor{violet!50} $A$ & 0.00\% & 0.14\% & 41.9\\ \hline
    \cellcolor{violet!50} $A$ & \cellcolor{violet!50} $A$ & \cellcolor{violet!50} $A$ & \cellcolor{orange!50} $E$ & 0.03\% & 0.43\% & 38.0\\ \hline
    \cellcolor{green!50} $C$ & \cellcolor{violet!50} $A$ & \cellcolor{violet!50} $A$ & \cellcolor{violet!50} $A$ & 0.00\% & 0.07\% & 35.2\\    
    \bottomrule[1pt]
  \end{tabular}
  \caption{Most frequent sub-sequences and $Z$-scores. All sequences in tables appear significantly above or below chance at $p<0.001$ (with Bonferroni correction).}
  \label{table:frequent_sequences_tests}
\end{table}



\subsection{Behavioral motifs}
A synthesis of the results presented above points to the emergence of several typical motifs, which shed light on the nature of co- contributor sequences in Wikipedia. In trying to interpret the meaning of the emerging patterns, we rely on qualitative analysis. Below, we describe the key patterns that emerge and explain their implications for our understanding of generative routines within peer-production.

\subsubsection{Solo contributor motifs}
\begin{table}[tb]
  \centering
  \begin{tabular}{c|c|c|c|l|r|r}
    \toprule[1pt]
    \multicolumn{4}{c|}{\textbf{Label}} & \textbf{Example} & \textbf{Count} & \textbf{Z-Score} \\
    1 & 2 & 3 & 4 & & \\
    \toprule[1pt]
	\cellcolor{violet!50} $A$ & \cellcolor{violet!50} $A$ & & & $U_1$-$U_1$-$U_1$ & 839 & 51.2\\ \hline
    \cellcolor{violet!50} $A$ & \cellcolor{violet!50} $A$ & \cellcolor{violet!50} $A$ & & $U_1$-$U_1$-$U_1$-$U_1$ & 487 & 130.8\\ \hline
    \cellcolor{violet!50} $A$ & \cellcolor{violet!50} $A$ & \cellcolor{orange!50} $E$ & & $U_1$-$U_1$-$U_1$-$U_2$ & 245 & 12.1\\ \hline
    \cellcolor{violet!50} $A$ & \cellcolor{violet!50} $A$ & \cellcolor{violet!50} $A$ & \cellcolor{violet!50} $A$ & $U_1$-$U_1$-$U_1$-$U_1$-$U_1$ & 325 & 323.5\\ \hline
	\cellcolor{violet!50} $A$ & \cellcolor{violet!50} $A$ & \cellcolor{violet!50} $A$ & \cellcolor{orange!50} $E$ & $U_1$-$U_1$-$U_1$-$U_1$-$U_2$ & 122 & 38.0\\
    \bottomrule[1pt]
  \end{tabular}
  \caption{Uninterrupted Same-contributor sequences, counts, and $Z$-scores. All sequences in tables appear significantly above or below chance at $p<0.001$ (with Bonferroni correction).}
  \label{table:uninterrupted_sequences_tests}
\end{table}

Much of the production work in Wikipedia is the result of a series of consecutive editing sessions made by the same contributor. Despite same-user editing session (label $A$) to make up only 8\% from all labels, many of the multi-label motifs include a series of $A$ labels. Commonly an $A$ editing session is the result of the collapse of several edits (on average spanning 3.15 minutes). It is interesting to note that even after collapsing immediate same-contributor edits onto editing sessions, we still observe in Table~\ref{table:uninterrupted_sequences_tests} an above-chance occurrence of sequences of label $A$.  Often, $A$ label editing sessions represent a quick reaction to a previous editing session (average time between is 2 days).

The literature on peer-production has paid little attention to the nature of co-contributor sequences. A recent work on open source software development projects suggests that (a) the overwhelming majority of work is accomplished with only a single programmer working on a task and (b) when tasks appear too large for an individual they are more likely to be deferred until they are easier, rather than being undertaken through structured teamwork~\cite{howison_collaboration_2014}. Our findings suggest that Wikipedia, too, may rely on a single contributor work sessions (as opposed to multi-contributor teamwork). This result stands in contrast to earlier accounts of Wikipedia which depicted co-production as a tight networked co-authorship.

\subsubsection{Reactive contributing motifs}
\begin{table}[tb]
  \centering
  \begin{tabular}{c|c|c|c|l|r|r}
    \toprule[1pt]
    \multicolumn{4}{c|}{\textbf{Label}} & \textbf{Example} & \textbf{Count} & \textbf{Z-Score} \\
    1 & 2 & 3 & 4 & & \\
    \toprule[1pt]
	\cellcolor{blue!50} $B$ & \cellcolor{blue!50} $B$ & & & $U_1$-$U_2$-$U_1$-$U_2$ & 393 & 25.7\\ \hline
    \cellcolor{blue!50} $B$ & \cellcolor{blue!50} $B$ & \cellcolor{blue!50} $B$ & & $U_1$-$U_2$-$U_1$-$U_2$-$U_1$ & 165 & 58.2\\ \hline
    \cellcolor{blue!50} $B$ & \cellcolor{blue!50} $B$ & \cellcolor{orange!50} $E$ & & $U_1$-$U_2$-$U_1$-$U_2$-$U_3$ & 167 & 10.1\\ \hline
    \cellcolor{orange!50} $E$ & \cellcolor{blue!50} $B$ & \cellcolor{blue!50} $B$ & & $U_2$-$U_1$-$U_2$-$U_1$ & 136 & 6.5\\ \hline
    \cellcolor{blue!50} $B$ & \cellcolor{blue!50} $B$ & \cellcolor{blue!50} $B$ & \cellcolor{blue!50} $B$ & $U_1$-$U_2$-$U_1$-$U_2$-$U_1$-$U_2$ & 83 & 121.9\\ \hline
    \cellcolor{blue!50} $B$ & \cellcolor{blue!50} $B$ & \cellcolor{blue!50} $B$ & \cellcolor{orange!50} $E$ & $U_1$-$U_2$-$U_1$-$U_2$-$U_1$-$U_3$ & 53 & 21.6\\ \hline
    \cellcolor{blue!50} $B$ & \cellcolor{blue!50} $B$ & \cellcolor{orange!50} $E$ & \cellcolor{orange!50} $E$ & $U_1$-$U_2$-$U_1$-$U_2$-$U_3$-$U_4$ & 100 & 6.5\\ 
    \bottomrule[1pt]
  \end{tabular}
  \caption{Reactive contributor sequences, counts, and $Z$-scores. All sequences in tables appear significantly above or below chance at $p<0.001$ (with Bonferroni correction).}
  \label{table:reactive_sequences_tests}
\end{table}
The second interesting pattern we observe represents back-and-forth editing sessions (``ping-pongs'') between active contributors. Two-contributor ping-pongs may occur in immediate sequence (\textit{e.g.}, $B$--$B$ motifs) or may be interrupted by a third user edit (\textit{e.g.}, $B$--$C$--$B$--$B$), and are characterized by a short duration between editing sessions (1.7 days for $B$ label, compared to the 9-day average across all labels). As illustrated in Table~\ref{table:reactive_sequences_tests}, such motifs occur significantly more than what is expected by chance alone.


Remarkably, despite the underlying assumption that large-scale technology-mediated online collaborations are the result of multiple contributors working together, the literature on Wikipedia – and more broadly, studies of online production communities – have rarely provided empirical accounts of co-production work that involves a restricted set of individuals working together. Interestingly, the one form of ping-pong that has been discussed in prior studies concerns destructive editing activities (\textit{e.g.}, edit wars between vandals and vandalism fighters)~\cite{kittur_he_2007, viegas_studying_2004, yasseri_dynamics_2012}.

A qualitative analysis of a selective set of ``ping-pong'' motifs shows that many of those motifs correspond to edit wars (in almost all cases, the vandal was a non-registered member, identified only by his IP address). For example, an analysis of the ``Flying Car'' article found that active-contributor sequences entailed vandalism-revert edit wars in August 2006 (4 vandalism edits, each followed by a revert), June 2007 (3 vandalism-revert pairs), and May 2008 (vandalism-revert-vandalism-revert). Label $B$ is characteristic of these ping-pongs and reflects an average large edit (over 1400 characters; more than double the average size of all labels), possibly reflecting `revert' type edits (which are the size of the entire article content). Nonetheless, we did also observe few-and-active contributor sequences that represent consecutive sessions by constructive activity. For example, we observed multiple periods of constructive ping-pongs for the ``Orange Revolution'' article: in February 2006, August 2006 and January 2010 (a series of sequential revisions by two contributors, where one contributor adds new content and the other reorganizes content on the page).

\subsubsection{Inactive contributor motifs}
\begin{table}[tb]
  \centering
  \begin{tabular}{c|c|c|c|l|r|r}
    \toprule[1pt]
    \multicolumn{4}{c|}{\textbf{Label}} & \textbf{Example} & \textbf{Count} & \textbf{Z-Score} \\
    1 & 2 & 3 & 4 & & \\
    \toprule[1pt]
	\cellcolor{orange!50} $E$ & \cellcolor{orange!50} $E$ & & & $U_1$-$U_2$ & 14,003 & 9.6\\ \hline
    \cellcolor{orange!50} $E$ & \cellcolor{orange!50} $E$ & \cellcolor{orange!50} $E$ & & $U_1$-$U_2$-$U_3$ & 10,469 & 20.1\\ \hline
    \cellcolor{orange!50} $E$ & \cellcolor{orange!50} $E$ & \cellcolor{orange!50} $E$ & \cellcolor{orange!50} $E$ & $U_1$-$U_2$-$U_3$-$U_4$ & 7,973 & 28.7\\
    \bottomrule[1pt]
  \end{tabular}
  \caption{Common inactive contributor sequences, counts, and $Z$-scores. All sequences in tables appear significantly above or below chance at $p<0.001$ (with Bonferroni correction).}
  \label{table:common_inactive_sequences}
\end{table}

\begin{table*}[tb]
  \centering
  \begin{tabular}{c|c|c|c|l|r|r}
    \toprule[1pt]
    \multicolumn{4}{c|}{\textbf{Label}} & \textbf{Example} & \textbf{Count} & \textbf{Z-Score} \\
    1 & 2 & 3 & 4 & & \\
    \toprule[1pt]
	\cellcolor{orange!50} $E$ & \cellcolor{violet!50} $A$ & & & $U_1$-$U_1$ & 815 & -18.2\\ \hline
    \cellcolor{orange!50} $E$ & \cellcolor{violet!50} $A$ & \cellcolor{orange!50} $E$ & & $U_1$-$U_1$-$U_2$ & 493 & -16.6\\ \hline
    \cellcolor{orange!50} $E$ & \cellcolor{orange!50} $E$ & \cellcolor{violet!50} $A$ & & $U_1$-$U_2$-$U_2$ & 565 & -14.3\\ \hline
    \cellcolor{yellow!50} $D$ & \cellcolor{orange!50} $E$ & \cellcolor{violet!50} $A$ & & $U_1$-$U_3$-$U_4$-$U_3$-$U_3$-$U_4$-$U_1$-$U_2$-$U_2$ & 107 & -6.8\\ \hline
    \cellcolor{violet!50} $A$ & \cellcolor{orange!50} $E$ & \cellcolor{violet!50} $A$ & & $U_1$-$U_1$-$U_2$-$U_2$ & 58 & -5.4\\ \hline
    \cellcolor{orange!50} $E$ & \cellcolor{violet!50} $A$ & \cellcolor{orange!50} $E$ & \cellcolor{orange!50} $E$ & $U_1$-$U_1$-$U_2$-$U_3$ & 336 & -13.4\\ \hline
    \cellcolor{orange!50} $E$ & \cellcolor{orange!50} $E$ & \cellcolor{violet!50} $A$ & \cellcolor{orange!50} $E$ & $U_1$-$U_2$-$U_2$-$U_3$ & 356 & -12.7\\ \hline
    \cellcolor{orange!50} $E$ & \cellcolor{orange!50} $E$ & \cellcolor{orange!50} $E$ & \cellcolor{violet!50} $A$ & $U_1$-$U_2$-$U_3$-$U_3$ & 431 & -9.8\\ \hline
    \cellcolor{yellow!50} $D$ & \cellcolor{orange!50} $E$ & \cellcolor{orange!50} $E$ & \cellcolor{violet!50} $A$ & $U_1$-$U_4$-$U_6$-$U_7$-$U_4$-$U_5$-$U_1$-$U_2$-$U_3$-$U_3$ & 56 & -7.0\\ \hline
    \cellcolor{yellow!50} $D$ & \cellcolor{orange!50} $E$ & \cellcolor{violet!50} $A$ & \cellcolor{orange!50} $E$ & $U_1$-$U_4$-$U_6$-$U_6$-$U_4$-$U_5$-$U_1$-$U_2$-$U_2$-$U_3$ & 61 & -6.6\\
    \bottomrule[1pt]
  \end{tabular}
  \caption{New-comer contributor non-sequential sequences, counts, and $Z$-scores. All sequences in tables appear significantly above or below chance at $p<0.001$ (with Bonferroni correction).}
  \label{table:newcomer_interrupted_sequences}
\end{table*}

The third pattern emerging from our analysis is that those editing the article for the first time tend to arrive consequently. Moreover, editing sessions represented by the label $E$ often correspond to a single edit: their editing sessions are not the result of collapsing multiple edits (their average duration is 0.75 minutes). Despite the high expected appearance for label $E$ (comprising close to 70\% of all labels), motifs representing a series of $E$ labels appear significantly more than anticipated by chance alone. Table~\ref{table:common_inactive_sequences} describes those sequences appearing significantly more than was is expected by chance alone, while Table~\ref{table:newcomer_interrupted_sequences} presents examples for the reverse effect –-- new editors on an article remain engaged –-- showing that it appears significantly below chance. 

We offer two alternative interpretations for these sequences. First, we proposed that sequences of first-time editors may represent dormant periods where ``owners'' do not attend to the article. This is implied by the lengthy times between the preceding editing session and the label $E$ session (over 11 days). Similarly, the duration leading to first-time-Wikipedia-edit (label $F$) is also long, over 6 days. Our second explanation suggests that there are sudden shocks (events internal to Wikipedia or external) that are attracting editing activities by first-time contributors. Earlier accounts of Wikipedia have demonstrated that articles about breaking news incidents exhibit high-tempo coordination dynamics, with structures and dynamics distinct from those observed among articles about non-breaking events~\cite{keegan_hot_2013}. 

In order to explore these bursts of activity by peripheral members, we performed a follow-on analysis seeking to identify the event that triggered those sequences. We studied Google searches for the keywords corresponding to articles’ titles, finding that often peaks in the usage of that particular keywords are associated with the start of $E$ sequences, implying that an external event (attracting the Google searches) has triggered the burst of activity by those first-time contributors. For example, Google searches for the Canadian actor Cameron Bright peeked in August 2009 (possibly associated with the anticipated release of the film ``The Twilight Saga: New Moon'' where he played a leading role), and at the same time the Wikipedia article for Cameron Bright saw a burst of activity by first-time arrivals to this article. 

\subsubsection{Distinctive motifs}
\begin{table}[tb]
  \centering
  \begin{tabular}{c|c|c|l|r|r}
    \toprule[1pt]
    \multicolumn{3}{c|}{\textbf{Label}} & \textbf{Example} & \textbf{Count} & \textbf{Z-Score} \\
    1 & 2 & 3 & & \\
    \toprule[1pt]
	\cellcolor{violet!50} $A$ & \cellcolor{orange!50} $E$ & & $U_1$-$U_1$-$U_2$ & 1,003 & -13.2\\ \hline
    \cellcolor{blue!50} $B$ & \cellcolor{orange!50} $E$ & & $U_1$-$U_2$-$U_1$-$U_3$ & 986 & -7.3\\ \hline
    \cellcolor{green!50} $C$ & \cellcolor{orange!50} $E$ & & $U_1$-$U_3$-$U_4$-$U_1$-$U_2$ & 791 & -4.1\\ \hline
    \cellcolor{orange!50} $E$ & \cellcolor{violet!50} $A$ & \cellcolor{orange!50} $E$ & $U_1$-$U_1$-$U_2$ & 493 & -16.6\\ \hline
    \cellcolor{violet!50} $A$ & \cellcolor{orange!50} $E$ & \cellcolor{orange!50} $E$ & $U_1$-$U_1$-$U_2$-$U_3$ & 639 & -12.0\\ \hline
    \cellcolor{blue!50} $B$ & \cellcolor{orange!50} $E$ & \cellcolor{orange!50} $E$ & $U_1$-$U_2$-$U_1$-$U_3$-$U_4$ & 612 & -7.8\\
    \bottomrule[1pt]
  \end{tabular}
  \caption{Unusual contributor sequences, counts, and $Z$-scores. All sequences in tables appear significantly above or below chance at $p<0.001$ (with Bonferroni correction).}
  \label{table:distinctive_sequences}
\end{table}

The forth pattern we observe in Table~\ref{table:distinctive_sequences} is that newcomers to the article (label $E$) are unlikely to contribute after an editing session made by a previously-active contributor (labels $A$, $B$, or $C$). The rare occurrence of $A$--$E$ and $C$--$E$ motifs implies that there are activity periods where insiders and ``owners'' of an article concentrate their effort. During those times, outsiders are unlikely to join the co-authoring effort. To the best of our knowledge, this particular motif has not been reported in prior studies of peer production.

\subsection{Case Study Discussion}
Our case study illustrates the potential of sequence analysis to illuminate deeper patterns within online knowledge collaborations using a mixture of qualitative and quantitative methods~\cite{small_how_2011}. We used a method to capture relative activity of editors, defined a set of sequential patterns, measured their frequency within a cross-sample of Wikipedia data, and tested the likelihood of observing patterns compared to chance and using the Bonferroni correction to increase confidence in the validity of the significance of the observed results. Having identified the prevalence of many quantitative patterns occurring significantly more than random, we employed a qualitative analysis to iteratively code the data, identifying different types of multi-label motifs. Grouping motifs that represent similar sequences characteristics, we were able to identify four distinct key patterns. Some of these patterns were already documented in Wikipedia (\textit{e.g.}, destructive ping-pongs), others reported for other peer-production projects but not in Wikipedia (\textit{e.g.}, co-production relying on a single contributor work sessions, as opposed to multi-contributor teamwork), and yet others have not yet been documented in Wikipedia or more broadly in peer-production (\textit{e.g.}, sequences of productive ping-pongs).

This case study is illustrative of the potential of a sustained research agenda that would employ sequence analysis methods originally developed in sociology, biology, and other fields, in order to analyze complex organizational processes within peer production communities and online knowledge collaborations~\cite{gaskin_toward_2014, yoo_digital_2012}. Event logs archiving the activities, artifacts, performers, and order of actions are common across socio-technical systems, but only a handful of analyses to date have explored the temporal or sequential patterns within these activity sequences. We outlined a preliminary set of approaches for defining different relationships within sequences based on varying levels of analysis. Having identified sequences from socio-technical event log data, a range of established methods from pattern mining, sequence similarity, and probabilistic analysis are available for quantitative modeling. 

\section{Sequence Analysis Research Agenda for CSCW}
While sociologists have championed the role of sequence data for understanding questions about the prevalence of patterns within sequences,  the independent predictors of sequence patterns, and the consequences of sequence patterns~\cite{abbott_primer_1990}, these research questions have made few in-roads to human-computer interaction (HCI) or computer-supported cooperative work (CSCW). Having described our proposed framework for sequence analysis of socio-technical trajectories and having illustrated its application through a case study, we now turn our attention to outlining a research agenda for the application of this framework to answer questions about the configuration, prevalence, and relationship of sequence data in the field of CSCW. We first discuss this research agenda at the methodological level, describing various types of research objectives that could be addressed through sequence analysis. We then proceed to review common CSCW research domains and settings and provide examples of how our framework could be applied across them.

\subsection{Relationships within Sequences}
Sequences are fundamentally relational constructs that connect a series of activities, artifacts, and performers together. We primarily define these sequences by a time order, where elements are made to be adjacent if they occur after each other. Instead of generating an artifact-oriented event log by sub-setting Wikipedia's entire event log by a single article's event log as we did in Table~\ref{table:event_log}, we could also subset Wikipedia's event log by a single user's contribution history. This user-oriented event log would reveal the sequence of actions of a single user constituting their editing history over time and across artifacts, documenting the accumulation of actions that ultimately define a social identity or a role~\cite{arazy_functional_2015, gleave_conceptual_2009}. Alternatively, sub-setting on a single edit type~\cite{kriplean_articulations_2008} could reveal distinctive sub-classes of work or the evolution of routines. Each of the sequences generated from these smaller subsets can then be used as a case and compared against each other in subsequent analyses. Thus a sequence of multi-dimensional events might be ``projected'' onto a single dimension. That is, a single contribution can be a member of different types of sequences depending on the unit of analysis employed at the performer (editor), artifact (article), or activity (edit type) levels, and these different sequences can be compared in turn.


\subsection{Recurring Patterns across Sequences}
Questions about patterns within sequences primarily ask whether there are typical sub-sequences.  How common or unusual is a pattern or motif? Do some actions reliably follow others? Can we interpret a cluster of highly-similar sequences as representing a class of behavior? Furthermore, the potential of extracting multiple sequences from event logs likewise invite questions about the alignments between these different sequences, for example: are changes in activity types coupled with changes in performers? Finally the use of probabilistic models allows us to analyze changes in the likelihood of state transitions between a series of adjacent sequences.
\newline

\subsection{Interpretation of Sequences}

Event log data in socio-technical systems like Wikipedia and GitHub enables a kind of archival ethnography (or information archaeology) by which the preceding and subsequent actions in the past can be re-read and re-interpreted, much as a contemporaneous user would have also interacted with the subject~\cite{christine_2000_virtual}. Sequence analysis methods can be used as a way of triangulating between nested approaches: ``zooming out'' from simple dyadic relationships to identify macro-social statistical patterns about similarity, and ``zooming in'' to narratively interpret quantitative outliers as the consequence of specific contexts, emergent patterns, or crucial turning points that the quantification overlooked ~\cite{small_how_2011}. Qualitative analysis is likewise crucial for identifying breaks in the chain of reasoning and threats to validity when translating data extracted from information systems through analytical modeling and transformed to the relevant analytical constructs under study~\cite{howison2011validity}.



\subsection{Antecedents of Sequenced Behavior}
The patterns observed in sequences are not random, but rather reflect path-dependencies underlying social processes such as collaboration and conflict. This class of questions can explore how other behavioral or psychological constructs of people or technological affordances of a technology influence the kinds of sequences that activities, artifacts, and performers subsequently become embedded within. For example, Wikipedia editors engaged in anti-vandalism work are likely to have very distinct sequential patterns in their contribution behavior compared to editors engaged in copy-writing. The organizational and external contexts also influences the kinds of sequences observed: high-tempo collaborations following crises may prompt different kinds of sequences than those following traditional, low-tempo collaborations~\cite{keegan_staying_2012}.

\subsection{Consequences of Sequenced Behavior}
Sequences may influence other outcomes observed in online knowledge collaborations. Research may explore how certain activity patterns influence individuals’ performance (\textit{e.g.}, administrative promotions) as well as organizational performance (\textit{e.g.}, the production of high-quality content). Investigating the predictors of article quality in Wikipedia is an active research area~\cite{arazy_measurability_2011, ransbotham_membership_2011} and future research could explore the relationships between particular sequences of activity and article quality. In addition, the organizational literature discusses ways in which joint involvement in routines could facilitate the formation of relationships~\cite{feldman_organizational_2002}, thus a possible area for exploring the relationship between co-creation sequences and the creation of interpersonal relationships.

\begin{table*}[tb]
  \centering
  \begin{tabular}{l|p{13cm}}
    \toprule[1pt]
    \textbf{Topic} & \textbf{Research Questions}\\
    \toprule[1pt]
    Collaborative filtering & \parbox{\textwidth}{
            \begin{itemize}
                \item How do users' rating of the same items vary based on the position of the items within\\ a sequence of users' other ratings? (\textbf{Recurring Pattern}) \vspace{-.5em}
                \item What are features that predict the beginning and end of a user's rating sessions? \\(\textbf{Antecedent}) \vspace{-.5em}
                \item How does the accuracy of recommendations using sequence similarity compare\\ to traditional collaborative filtering methods? (\textbf{Consequence})
            \end{itemize} }\\\midrule
            
    Online social networks & \parbox{\textwidth}{
            \begin{itemize}
                \item Are there similarities in the sequences of users' creating and deleting relationships?\\(\textbf{Relationships}) \vspace{-.5em}
                \item Do users' information seeking sequences differ by demographics, skill, personality, or\\relationship strength? (\textbf{Antecedent}) \vspace{-.5em}
                \item How do variations in the sequence of items presented in a stream or feed influence user\\engagement? (\textbf{Consequence})
            \end{itemize} }\\\midrule
            
    Crowdsourcing & \parbox{\textwidth}{
            \begin{itemize}
                \item Are there sequences in a worker's early history that predicts their subsequent level of\\commitment or quality? (\textbf{Recurring Pattern}) \vspace{-.5em}
                \item Can the sequences of HITs completed by multiple workers reveal shared latent interests?\\ (\textbf{Interpretation}) \vspace{-.5em}
                \item How does HIT quality vary in relation to the task demands of prior tasks? (\textbf{Consequence}) 
            \end{itemize} }\\\midrule
            
    Online education & \parbox{\textwidth}{
            \begin{itemize}
                \item How do specific teaching elements disrupt the flow of successfully completing course\\content? (\textbf{Recurring Pattern}) \vspace{-.5em}
                \item What do clusters of course participation sequences reveal about students' interests and\\backgrounds? (\textbf{Antecedent}) \vspace{-.5em}
                \item Does the sequence of participation in prior courses influence subsequent course\\engagement and performance? (\textbf{Consequence})
            \end{itemize} }\\\midrule
            
    Crisis informatics & \parbox{\textwidth}{
            \begin{itemize}
                \item What are common behavioral sequences taken by users in the immediate aftermath of\\an event? (\textbf{Recurring Pattern}) \vspace{-.5em}
                \item How do users adopt social roles and employ routines in response to the sequence\\of new information about an event? (\textbf{Interpretation}) \vspace{-.5em}
                \item How were pre-existing routines and behaviors disrupted by an event and how long did\\this disruption persist? (\textbf{Consequence})
            \end{itemize} }\\\midrule
            
    Citizen science & \parbox{\textwidth}{
            \begin{itemize}
                \item To what extent do contributors’ motivations and attitudes determine their future\\participation patterns? (\textbf{Antecedents}) \vspace{-.5em}
                \item What are the sequences of activities that predict one’s commitment, formal positions\\taken, and sustained participation? (\textbf{Consequences}) \vspace{-.5em}
                \item What do the sequences of feedback (ratings, comments) around a reported observation\\tell us about the reliability of this observation? (\textbf{Interpretation})
            \end{itemize} }\\
    \bottomrule[1pt]
  \end{tabular}
  \caption{Examples of research questions for extending sequence analysis methods into other CSCW topics.}
  \label{table:CSCW_implications}
\end{table*}

\subsection{Implications for CSCW Research}
Our framing and case study emphasized the relevance of sequence analysis methods for understanding online knowledge collaborations such as peer-production (and in particular, Wikipedia). However, these methods also have clear applications to other CSCW domains. Table 11 identifies six established and emerging CSCW sub-topics and provides examples for research questions that could be addressed by employing sequence analysis approaches. We illustrate each CSCW sub-topic with three research questions, as well as map each of these research questions back to one of the research objectives described above. This table is not intended to exhaust all of the potential sub-topics that exist within CSCW, but rather to illustrate the potential for the proposed approach across a variety of areas that have been at the center of CSCW research:  collaborative filtering, online social networks, crowdsourcing, online education, crisis informatics, and citizen science, among many other domains. The analysis summarized in Table~\ref{table:CSCW_implications} outlines how sequence analysis could extend our understanding of socio-technical behavior across diverse CSCW domains.




\section{Conclusion}
Event log data in socio-technical systems like Wikipedia encode a rich variety of content and meta-data about who is contributing what, where, when, and how. Our framework for analyzing behavioral sequences in event log data intends to answer questions pertaining to the structure and dynamics of online knowledge collaboration. Our research framework adapts a generalizable definition of event logs, and outlines the process of identifying, schematizing, analyzing, and interpreting sequences of activities, artifacts, or performers that co-occur. It is important to emphasize that interest in studying the temporal dynamics within online collaboration is not new; however, prior studies investigating dynamics within peer-production tended to “collapse” and ``aggregate'' data, thus losing important details regarding temporal sequences. Sequence analysis methods are able to mitigate these concerns and provide a richer description of the temporal dynamics underlying socio-technical systems. We outlined a step-by-step procedure for employing sequence analysis in a variety of CSCW domains and in demonstrating the validity of this approach through a case study that explored the prevalence and statistical significance of contributors' activity sequences in the context of Wikipedia.


This case study examining Wikipedia editors' sequential edits to articles represents an implementation of the ``grammatical'' models proposed by theorists of ``practice'' in organizational studies. Much as language permits variation in form to convey similar meanings, organizations may support variations in patterns of actions to achieve similar goals. The prevalence (or absence) of repeated patterns of behavior in both structural and performative processes has been a cornerstone of theoretical developments in the areas of organizational routines and practice theory. However, empirically testing these grammatical models has been limited by the difficulty of obtaining data from ethnographic field studies. To date, there is very little empirical support for these grammatical theories of organizational routines. Socio-technical event log data has the variability, breadth, and granularity to adapt grammatical models of organizational routines to understand the variation, selection, and retention of new practices and patterns of actions in technology-mediated social participation. By schematizing contributors’ activity histories relative to each other, we were able to identify patterns of action occurring significantly above (or below) chance. Findings from our empirical case study provide support for Feldman's idea of performativity in organizational routines, whereby people adopt or abandon routines in response to others’ actions.


As a methodological contribution, this research agenda is a call to extend methods for examining and interpreting the relationships, patterns, antecedents, and consequences of sequences in event log data across socio-technical systems. While underutilized within CSCW, methods for analyzing, aligning, and modeling sequences are well-developed in fields like bio-informatics, thus inviting potentially compelling analogies for future theoretical elaboration. Far from being a framework applicable only to peer production or online knowledge collaborations, sequence analysis methods have substantial potential to become a first-class method to answer research questions about event log data. Consider natural language processing or network science: both provide researchers with a set of tools drawn from to extract features and structured information from complex, multidimensional data; both have long, interdisciplinary histories; and both have been adapted to support mixed methods research. We foresee a future where scholars studying online group processes could draw from a portfolio of techniques that includes sequence approaches alongside CSCW's other mixed methods like content or network analysis.

Although developed as a quantitative method, sequence analysis approaches have very high potential for supporting mixed methods inquiry in the context of event log analysis. Results from sequence analysis can identify common patterns or outliers, which invite closer scrutiny. Because sequences involve actors and artifacts with attributes like status or quality (respectively), these co-variates allow comparison of different kinds of data measuring the same phenomena. In addition to juxtaposing across distinct kinds of data, these data can also be modeled in relation to each other as predictors or consequences of other processes. In effect, sequence analysis methods can provide a set of second-order behavioral attributes that are conceptually and theoretically distinct from first-order attributes like counts or intensities. Notwithstanding the merits of methods such as content analysis of textual data or network analysis of dyadic interactions, we maintain that integrating first- and second-order behavior features, opens up a whole new set of ways to compliment and confirm other approaches for contextualizing, triangulating, and confirming socio-technical behaviors.

\section{Acknowledgments}
We would like to thank Darren Gergle and the reviewers for their valuable feedback on an earlier version of this paper, Adam Balila for assistance with data extraction and organization, Deborah Keegan for copy-editing, and the Wikimedia Foundation and Wikipedians who support it for making these data available for research. This work was partially supported by SSHRC Insight Grant 435-2013-0624.

\balance

\bibliographystyle{acm-sigchi}
\bibliography{paper}


\begin{thebibliography}{00}


\ifx \showCODEN    \undefined \def \showCODEN     #1{\unskip}     \fi
\ifx \showDOI      \undefined \def \showDOI       #1{{\tt DOI:}\penalty0{#1}\ }
  \fi
\ifx \showISBNx    \undefined \def \showISBNx     #1{\unskip}     \fi
\ifx \showISBNxiii \undefined \def \showISBNxiii  #1{\unskip}     \fi
\ifx \showISSN     \undefined \def \showISSN      #1{\unskip}     \fi
\ifx \showLCCN     \undefined \def \showLCCN      #1{\unskip}     \fi
\ifx \shownote     \undefined \def \shownote      #1{#1}          \fi
\ifx \showarticletitle \undefined \def \showarticletitle #1{#1}   \fi
\ifx \showURL      \undefined \def \showURL       #1{#1}          \fi

\bibitem{abbott_primer_1990}
{Andrew Abbott}. 1990.
\newblock \showarticletitle{A primer on sequence methods}.
\newblock {\em Organization Science\/} {1}, 4 (1990), 375–392.
\newblock


\bibitem{abbott_sequence_1995}
{Andrew Abbott}. 1995.
\newblock \showarticletitle{Sequence analysis: new methods for old ideas}.
\newblock {\em Annual Review of sociology\/} (1995), 93–113.
\newblock


\bibitem{agarwal2008digitalsocial}
{Ritu Agarwal}, {Anil~K. Gupta}, {and} {Robert Kraut}. 2008.
\newblock \showarticletitle{Editorial Overview--The Interplay Between Digital
  and Social Networks}.
\newblock {\em Information Systems Research\/} {19}, 3 (2008), 243--252.
\newblock
\showDOI{%
\url{http://dx.doi.org/10.1287/isre.1080.0200}}


\bibitem{arazy_sustainability_2010}
{Ofer Arazy} {and} {Arie Croitoru}. 2010.
\newblock \showarticletitle{The {Sustainability} of {Corporate} {Wikis}: {A}
  {Time}-series {Analysis} of {Activity} {Patterns}}.
\newblock {\em ACM Trans. Management Information Systems\/} {1}, 1 (2010),
  6:1--6:24.
\newblock
\showDOI{%
\url{http://dx.doi.org/10.1145/1877725.1877731}}


\bibitem{arazy_measurability_2011}
{Ofer Arazy} {and} {Rick Kopak}. 2011.
\newblock \showarticletitle{On the measurability of information quality}.
\newblock {\em Journal of the American Society for Information Science and
  Technology\/} {62}, 1 (2011), 89--99.
\newblock
\showISSN{1532-2890}
\showDOI{%
\url{http://dx.doi.org/10.1002/asi.21447}}


\bibitem{arazy_information_2011}
{Ofer Arazy}, {Oded Nov}, {Raymond Patterson}, {and} {Lisa Yeo}. 2011.
\newblock \showarticletitle{Information {Quality} in {Wikipedia}: {The}
  {Effects} of {Group} {Composition} and {Task} {Conflict}}.
\newblock {\em Journal of Management Information Systems\/} {27}, 4 (2011),
  71--98.
\newblock
\showISSN{0742-1222}
\showDOI{%
\url{http://dx.doi.org/10.2753/MIS0742-1222270403}}


\bibitem{arazy_functional_2015}
{Ofer Arazy}, {Felipe Ortega}, {Oded Nov}, {Lisa Yeo}, {and} {Adam Balila}.
  2015.
\newblock \showarticletitle{Functional Roles and Career Paths in {Wikipedia}}.
  In {\em Proc. CSCW 2015}. ACM, 1092--1105.
\newblock
\showISBNx{978-1-4503-2922-4}
\showDOI{%
\url{http://dx.doi.org/10.1145/2675133.2675257}}


\bibitem{becker_handbook_2008}
{Markus~C Becker}. 2008.
\newblock \showarticletitle{The past, present, and future of organizational
  routines}.
\newblock In {\em Handbook of Organizational Routines}. Edward Elgar
  Publishing.
\newblock


\bibitem{benjamini_controlling_1995}
{Yoav Benjamini} {and} {Yosef Hochberg}. 1995.
\newblock \showarticletitle{Controlling the {False} {Discovery} {Rate}: {A}
  {Practical} and {Powerful} {Approach} to {Multiple} {Testing}}.
\newblock {\em Journal of the Royal Statistical Society. Series B
  (Methodological)\/} {57}, 1 (1995), 289--300.
\newblock


\bibitem{Benkler2006wealth}
{Yochai Benkler}. 2006.
\newblock {\em The Wealth of Networks: How Social Production Transforms Markets
  and Freedom}.
\newblock Yale University Press, New Haven, CT, USA.
\newblock
\showISBNx{0300110561}


\bibitem{berndt_using_1994}
{Donald~J Berndt} {and} {James Clifford}. 1994.
\newblock \showarticletitle{Using {Dynamic} {Time} {Warping} to {Find}
  {Patterns} in {Time} {Series}.}. In {\em Proc. {KDD} 1994}. AAAI, 359--370.
\newblock


\bibitem{bryant_becoming_2005}
{Susan~L. Bryant}, {Andrea Forte}, {and} {Amy Bruckman}. 2005.
\newblock \showarticletitle{Becoming {Wikipedian}: {Transformation} of
  {Participation} in a {Collaborative} {Online} {Encyclopedia}}. In {\em Proc.
  {GROUP} 2005} {\em ({GROUP} '05)}. ACM, 1--10.
\newblock
\showDOI{%
\url{http://dx.doi.org/10.1145/1099203.1099205}}


\bibitem{crowston_freelibre_2008}
{Kevin Crowston}, {Kangning Wei}, {James Howison}, {and} {Andrea Wiggins}.
  2008.
\newblock \showarticletitle{Free/{Libre} {Open}-source {Software}
  {Development}: {What} {We} {Know} and {What} {We} {Do} {Not} {Know}}.
\newblock {\em ACM Computing Survey\/} {44}, 2 (2008), 7:1--7:35.
\newblock
\showDOI{%
\url{http://dx.doi.org/10.1145/2089125.2089127}}


\bibitem{de_souza_seeking_2005}
{Cleidson De~Souza}, {Jon Froehlich}, {and} {Paul Dourish}. 2005.
\newblock \showarticletitle{Seeking the source: software source code as a
  social and technical artifact}. In {\em Proc. GROUP 2005}. 197–206.
\newblock
\showURL{%
\url{http://dl.acm.org/citation.cfm?id=1099239}}


\bibitem{dempster_maximum_1977}
{A.~P. Dempster}, {N.~M. Laird}, {and} {D.~B. Rubin}. 1977.
\newblock \showarticletitle{Maximum {Likelihood} from {Incomplete} {Data} via
  the {EM} {Algorithm}}.
\newblock {\em Journal of the Royal Statistical Society. Series B
  (Methodological)\/} {39}, 1 (1977), 1--38.
\newblock


\bibitem{dijkstra_measuring_1995}
{Wil Dijkstra} {and} {Toon Taris}. 1995.
\newblock \showarticletitle{Measuring the {Agreement} between {Sequences}}.
\newblock {\em Sociological Methods \& Research\/} {24}, 2 (1995), 214--231.
\newblock


\bibitem{ducheneaut_2005_socialization}
{Nicolas Ducheneaut}. 2005.
\newblock \showarticletitle{Socialization in an Open Source Software Community:
  A Socio-Technical Analysis}.
\newblock {\em Computer Supported Cooperative Work (CSCW)\/} {14}, 4 (2005),
  323--368.
\newblock
\showISSN{0925-9724}
\showDOI{%
\url{http://dx.doi.org/10.1007/s10606-005-9000-1}}


\bibitem{dunne_graphtrail_2012}
{Cody Dunne}, {Nathalie Henry~Riche}, {Bongshin Lee}, {Ronald Metoyer}, {and}
  {George Robertson}. 2012.
\newblock \showarticletitle{GraphTrail: analyzing large multivariate,
  heterogeneous networks while supporting exploration history}. In {\em Proc.
  CHI 2012}. ACM, 1663--1672.
\newblock
\showISBNx{978-1-4503-1015-4}
\showDOI{%
\url{http://dx.doi.org/10.1145/2207676.2208293}}


\bibitem{faraj2011knowledge}
{Samer Faraj}, {Sirkka~L Jarvenpaa}, {and} {Ann Majchrzak}. 2011.
\newblock \showarticletitle{Knowledge collaboration in online communities}.
\newblock {\em Organization Science\/} {22}, 5 (2011), 1224--1239.
\newblock


\bibitem{feldman_reconceptualizing_2003}
{MS Feldman} {and} {BT Pentland}. 2003.
\newblock \showarticletitle{Reconceptualizing Organizational Routines as a
  Source of Flexibility and Change}.
\newblock {\em Administrative Science Quarterly\/}  {48} (2003), 94--121.
\newblock
\newblock
\shownote{1.}


\bibitem{feldman_organizational_2000}
{Martha~S. Feldman}. 2000.
\newblock \showarticletitle{Organizational Routines as a Source of Continuous
  Change}.
\newblock {\em Organization Science\/} {11}, 6 (2000), 611--629.
\newblock
\showISSN{1047-7039}
\showDOI{%
\url{http://dx.doi.org/10.1287/orsc.11.6.611.12529}}


\bibitem{feldman_organizational_2002}
{Martha~S. Feldman} {and} {Anat Rafaeli}. 2002.
\newblock \showarticletitle{Organizational {Routines} as {Sources} of
  {Connections} and {Understandings}}.
\newblock {\em Journal of Management Studies\/} {39}, 3 (2002), 309--331.
\newblock
\showISSN{1467-6486}
\showDOI{%
\url{http://dx.doi.org/10.1111/1467-6486.00294}}


\bibitem{fisher2004everyday}
{Danyel Fisher} {and} {Paul Dourish}. 2004.
\newblock \showarticletitle{Social and temporal structures in everyday
  collaboration}. In {\em Proc. CHI 2004}. ACM, 551--558.
\newblock
\showISBNx{1-58113-702-8}
\showDOI{%
\url{http://dx.doi.org/10.1145/985692.985762}}


\bibitem{forte_decentralization_2009}
{Andrea Forte}, {Vanesa Larco}, {and} {Amy Bruckman}. 2009.
\newblock \showarticletitle{Decentralization in {Wikipedia} {Governance}}.
\newblock {\em Journal of Management Information Systems\/} {26}, 1 (2009),
  49--72.
\newblock
\showDOI{%
\url{http://dx.doi.org/10.2753/MIS0742-1222260103}}


\bibitem{gaskin_toward_2014}
{James Gaskin}, {Nicholas Berente}, {Kalle Lyytinen}, {and} {{Youngjin Yoo}}.
  2014.
\newblock \showarticletitle{Toward Generalizable Sociomaterial Inquiry: {A}
  Computational Approach for Zooming in and Out of Sociomaterial Routines}.
\newblock {\em MIS Quarterly\/} {38}, 3 (2014), 849--A12.
\newblock
\showISSN{02767783}


\bibitem{geiger2013sessions}
{R.~Stuart Geiger} {and} {Aaron Halfaker}. 2013.
\newblock \showarticletitle{Using edit sessions to measure participation in
  wikipedia}. In {\em Proc. CSCW 2013}. ACM, 861--870.
\newblock
\showISBNx{978-1-4503-1331-5}
\showDOI{%
\url{http://dx.doi.org/10.1145/2441776.2441873}}


\bibitem{gibson_taking_2005}
{David~R. Gibson}. 2005.
\newblock \showarticletitle{Taking Turns and Talking Ties: Networks and
  Conversational Interaction}.
\newblock {\it Amer. J. Sociology} {110}, 6 (2005), 1561–1597.
\newblock
\showURL{%
\url{http://www.jstor.org/stable/10.1086/428689}}


\bibitem{gittell_coordination_2004}
{Jody~Hoffer Gittell} {and} {Leigh Weiss}. 2004.
\newblock \showarticletitle{Coordination Networks Within and Across
  Organizations: {A} Multi-level Framework}.
\newblock {\em Journal of Management Studies\/} {41}, 1 (2004), 127--153.
\newblock
\showISSN{1467-6486}
\showDOI{%
\url{http://dx.doi.org/10.1111/j.1467-6486.2004.00424.x}}


\bibitem{gleave_conceptual_2009}
{Eric Gleave}, {Howard~T Welser}, {Thomas~M Lento}, {and} {Marc~A Smith}. 2009.
\newblock \showarticletitle{A conceptual and operational definition of'social
  role'in online community}. In {\em Proc. HICSS 2009}. IEEE, 1--11.
\newblock


\bibitem{goggins_social_2010}
{S.P. Goggins}, {M. Gallagher}, {J. Laffey}, {and} {C. Amelung}. 2010.
\newblock \showarticletitle{Social {Intelligence} in {Completely} {Online}
  {Groups} - {Toward} {Social} {Prosthetics} from {Log} {Data} {Analysis} and
  {Transformation}}. In {\em Proc. {SocialCom}}. IEEE, 500--507.
\newblock
\showDOI{%
\url{http://dx.doi.org/10.1109/SocialCom.2010.79}}


\bibitem{goggins_connecting_2014}
{Sean Goggins} {and} {Eva Petakovic}. 2014.
\newblock \showarticletitle{Connecting {Theory} to {Social} {Technology}
  {Platforms} {A} {Framework} for {Measuring} {Influence} in {Context}}.
\newblock {\em American Behavioral Scientist\/} {58}, 10 (2014), 1376--1392.
\newblock
\showDOI{%
\url{http://dx.doi.org/10.1177/0002764214527093}}


\bibitem{han_frequent_2007}
{Jiawei Han}, {Hong Cheng}, {Xin Dong}, {and} {Xifeng Yan}. 2007.
\newblock \showarticletitle{Frequent pattern mining: current status and future
  directions}.
\newblock {\em Data Mining and Knowledge Discovery\/} {15}, 1 (2007), 55--86.
\newblock


\bibitem{christine_2000_virtual}
{Christine Hine}. 2000.
\newblock {\em Virtual ethnography}.
\newblock Sage.
\newblock


\bibitem{howison_collaboration_2014}
{James Howison} {and} {Kevin Crowston}. 2014.
\newblock \showarticletitle{Collaboration {Through} {Open} {Superposition}: {A}
  {Theory} of the {Open} {Source} {Way}}.
\newblock {\em MIS Quarterly\/} {38}, 1 (2014), 29--A9.
\newblock


\bibitem{howison2011validity}
{James Howison} {and} {Andrea Wiggins}. 2011.
\newblock \showarticletitle{Validity Issues in the Use of Social Network
  Analysis with Digital Trace Data}.
\newblock {\em Journal of the Association for Information Systems\/} {12}, 12
  (2011), 767--797.
\newblock


\bibitem{jahnke_dynamics_2010}
{Isa Jahnke}. 2010.
\newblock \showarticletitle{Dynamics of social roles in a knowledge management
  community}.
\newblock {\em Computers in Human Behavior\/} {26}, 4 (2010), 533--546.
\newblock
\showDOI{%
\url{http://dx.doi.org/10.1016/j.chb.2009.08.010}}


\bibitem{jurgens2012temporal}
{David Jurgens} {and} {Tsai-Ching Lu}. 2012.
\newblock \showarticletitle{Temporal Motifs Reveal the Dynamics of Editor
  Interactions in Wikipedia.}. In {\em Proc. ICWSM 2012}. AAAI, 162--169.
\newblock


\bibitem{kane_emergent_2014}
{Gerald~C. Kane}, {Jeremiah Johnson}, {and} {Ann Majchrzak}. 2014.
\newblock \showarticletitle{Emergent {Life} {Cycle}: {The} {Tension} {Between}
  {Knowledge} {Change} and {Knowledge} {Retention} in {Open} {Online}
  {Coproduction} {Communities}}.
\newblock {\em Management Science\/} {60}, 12 (2014), 3026--3048.
\newblock
\showDOI{%
\url{http://dx.doi.org/10.1287/mnsc.2013.1855}}


\bibitem{keegan_staying_2012}
{Brian Keegan}, {Darren Gergle}, {and} {Noshir Contractor}. 2012.
\newblock \showarticletitle{Staying in the loop: structure and dynamics of
  {Wikipedia}'s breaking news collaborations}. In {\em Proc. WikiSym 2012}.
  ACM, 1:1--1:10.
\newblock
\showDOI{%
\url{http://dx.doi.org/10.1145/2462932.2462934}}


\bibitem{keegan_hot_2013}
{Brian Keegan}, {Darren Gergle}, {and} {Noshir Contractor}. 2013.
\newblock \showarticletitle{Hot {Off} the {Wiki} {Structures} and {Dynamics} of
  {Wikipedia}’s {Coverage} of {Breaking} {News} {Events}}.
\newblock {\em American Behavioral Scientist\/} {57}, 5 (2013), 595--622.
\newblock
\showDOI{%
\url{http://dx.doi.org/10.1177/0002764212469367}}


\bibitem{kittur_he_2007}
{Aniket Kittur}, {Bongwon Suh}, {Bryan~A. Pendleton}, {and} {Ed~H. Chi}. 2007.
\newblock \showarticletitle{He {Says}, {She} {Says}: {Conflict} and
  {Coordination} in {Wikipedia}}. In {\em Proc. CHI 2007}. ACM, New York, NY,
  USA, 453--462.
\newblock
\showDOI{%
\url{http://dx.doi.org/10.1145/1240624.1240698}}


\bibitem{kriplean_articulations_2008}
{Travis Kriplean}, {Ivan Beschastnikh}, {and} {David~W. McDonald}. 2008.
\newblock \showarticletitle{Articulations of Wikiwork: {Uncovering} Valued Work
  in {Wikipedia} Through Barnstars}. In {\em Proc. CSCW 2008}. ACM, 47--56.
\newblock
\showDOI{%
\url{http://dx.doi.org/10.1145/1460563.1460573}}


\bibitem{lafferty_conditional_2001}
{John~D. Lafferty}, {Andrew McCallum}, {and} {Fernando C.~N. Pereira}. 2001.
\newblock \showarticletitle{Conditional {Random} {Fields}: {Probabilistic}
  {Models} for {Segmenting} and {Labeling} {Sequence} {Data}}. In {\em Proc.
  ICML 2001}. Morgan Kaufmann Publishers Inc., 282--289.
\newblock


\bibitem{lakhani_how_2003}
{Karim~R Lakhani} {and} {Eric von Hippel}. 2003.
\newblock \showarticletitle{How open source software works: “free”
  user-to-user assistance}.
\newblock {\em Research Policy\/} {32}, 6 (2003), 923--943.
\newblock
\showDOI{%
\url{http://dx.doi.org/10.1016/S0048-7333(02)00095-1}}


\bibitem{lazer2009life}
{David Lazer}, {Alex~Sandy Pentland}, {Lada Adamic}, {Sinan Aral},
  {Albert~Laszlo Barabasi}, {Devon Brewer}, {Nicholas Christakis}, {Noshir
  Contractor}, {James Fowler}, {Myron Gutmann}, {and} {others}. 2009.
\newblock \showarticletitle{Life in the network: the coming age of
  computational social science}.
\newblock {\em Science\/} {323}, 5915 (2009), 721.
\newblock


\bibitem{lin_experiencing_2007}
{Jessica Lin}, {Eamonn Keogh}, {Wei Li}, {and} {Stefano Lonardi}. 2007.
\newblock \showarticletitle{Experiencing {SAX}: a novel symbolic representation
  of time series}.
\newblock {\em Data Mining and Knowledge Discovery\/} {15}, 2 (2007), 107--144.
\newblock


\bibitem{malone_interdisciplinary_1994}
{{TW} Malone} {and} {K Crowston}. 1994.
\newblock \showarticletitle{The interdisciplinary study of coordination}.
\newblock {\it Comput. Surveys}  {26} (1994), 119.
\newblock
\newblock
\shownote{1.}


\bibitem{nakakoji_evolution_2002}
{Kumiyo Nakakoji}, {Yasuhiro Yamamoto}, {Yoshiyuki Nishinaka}, {Kouichi
  Kishida}, {and} {Yunwen Ye}. 2002.
\newblock \showarticletitle{Evolution patterns of open-source software systems
  and communities}. In {\em Proceedings of the international workshop on
  Principles of software evolution}. 76–85.
\newblock
\showURL{%
\url{http://dl.acm.org/citation.cfm?id=512055}}


\bibitem{nelson_winter._2005}
{Richard~R Nelson} {and} {G Sidney}. 2005.
\newblock \showarticletitle{Winter. 1982. {An} evolutionary theory of economic
  change}.
\newblock {\em Cambridge: Belknap\/} (2005).
\newblock


\bibitem{olson_paying_2010}
{J.F. Olson}, {J. Howison}, {and} {K.M. Carley}. 2010.
\newblock \showarticletitle{Paying {Attention} to {Each} {Other} in {Visible}
  {Work} {Communities}: {Modeling} {Bursty} {Systems} of {Multiple} {Activity}
  {Streams}}. In {\em Proc. SocialCom 2010}. 276--281.
\newblock
\showDOI{%
\url{http://dx.doi.org/10.1109/SocialCom.2010.46}}


\bibitem{pentland_grammatical_1995}
{Brian~T. Pentland}. 1995.
\newblock \showarticletitle{Grammatical Models of Organizational Processes}.
\newblock {\em Organization Science\/} {6}, 5 (1995), 541--556.
\newblock
\showISSN{10477039}


\bibitem{pentland_narrative_2007}
{Brian~T. Pentland} {and} {Martha~S. Feldman}. 2007.
\newblock \showarticletitle{Narrative networks: Patterns of technology and
  organization}.
\newblock {\em Organization Science\/} {18}, 5 (2007), 781–795.
\newblock


\bibitem{pentland_designing_2008}
{Brian~T. Pentland} {and} {Martha~S. Feldman}. 2008.
\newblock \showarticletitle{Designing routines: On the folly of designing
  artifacts, while hoping for patterns of action}.
\newblock {\em Information and Organization\/} {18}, 4 (2008), 235--250.
\newblock
\showDOI{%
\url{http://dx.doi.org/10.1016/j.infoandorg.2008.08.001}}


\bibitem{pentland_organizational_1994}
{Brian~T. Pentland} {and} {Henry~H. Rueter}. 1994.
\newblock \showarticletitle{Organizational Routines as Grammars of Action}.
\newblock {\em Administrative Science Quarterly\/} {39}, 3 (1994), 484--510.
\newblock


\bibitem{ransbotham_membership_2011}
{Sam Ransbotham} {and} {Gerald~C. Kane}. 2011.
\newblock \showarticletitle{Membership {Turnover} and {Collaboration} {Success}
  in {Online} {Communities}: {Explaining} {Rises} and {Falls} from {Grace} in
  {Wikipedia}}.
\newblock {\em MIS Quarterly\/} {35}, 3 (2011), 613--627.
\newblock


\bibitem{salamin_introduction_2011}
{Hugues Salamin} {and} {Alessandro Vinciarelli}. 2011.
\newblock \showarticletitle{Introduction to {Sequence} {Analysis} for {Human}
  {Behavior} {Understanding}}.
\newblock In {\em Computer {Analysis} of {Human} {Behavior}}, {Albert~Ali
  Salah} {and} {Theo Gevers} (Eds.). Springer London, 21--40.
\newblock


\bibitem{shelly_sequences_1997}
{R.~K. Shelly}. 1997.
\newblock \showarticletitle{Sequences and Cycles in Social Interaction}.
\newblock {\em Small Group Research\/} {28}, 3 (1997), 333--356.
\newblock


\bibitem{small_how_2011}
{Mario~Luis Small}. 2011.
\newblock \showarticletitle{How to {Conduct} a {Mixed} {Methods} {Study}:
  {Recent} {Trends} in a {Rapidly} {Growing} {Literature}}.
\newblock {\em Annual Review of Sociology\/} {37}, 1 (2011), 57--86.
\newblock
\showDOI{%
\url{http://dx.doi.org/10.1146/annurev.soc.012809.102657}}


\bibitem{smith_visualization_2001}
{Marc~A. Smith} {and} {Andrew~T. Fiore}. 2001.
\newblock \showarticletitle{Visualization components for persistent
  conversations}. In {\em Proc. CHI 2001}. ACM, 136--143.
\newblock
\showISBNx{1-58113-327-8}
\showDOI{%
\url{http://dx.doi.org/10.1145/365024.365073}}


\bibitem{sundararajan_information_2013}
{Arun Sundararajan}, {Foster Provost}, {Gal Oestreicher-Singer}, {and} {Sinan
  Aral}. 2013.
\newblock \showarticletitle{Information in Digital, Economic, and Social
  Networks}.
\newblock {\em Information Systems Research\/} {24}, 4 (2013), 883--905.
\newblock
\showISSN{1047-7047, 1526-5536}
\showDOI{%
\url{http://dx.doi.org/10.1287/isre.1120.0472}}


\bibitem{tsay_influence_2014}
{Jason Tsay}, {Laura Dabbish}, {and} {James Herbsleb}. 2014.
\newblock \showarticletitle{Influence of {Social} and {Technical} {Factors} for
  {Evaluating} {Contribution} in {GitHub}}. In {\em Proc. ICSE 2014}. ACM,
  356--366.
\newblock
\showDOI{%
\url{http://dx.doi.org/10.1145/2568225.2568315}}


\bibitem{tushman_information_1978}
{Michael~L. Tushman} {and} {David~A. Nadler}. 1978.
\newblock \showarticletitle{Information Processing as an Integrating Concept in
  Organizational Design}.
\newblock {\em The Academy of Management Review\/} {3}, 3 (1978), 613--624.
\newblock
\showISSN{0363-7425}
\showDOI{%
\url{http://dx.doi.org/10.2307/257550}}


\bibitem{van2005discovering}
{Wil~MP van~der Aalst}, {Hajo~A Reijers}, {and} {Minseok Song}. 2005.
\newblock \showarticletitle{Discovering social networks from event logs}.
\newblock {\em Computer Supported Cooperative Work\/} {14}, 6 (2005), 549--593.
\newblock


\bibitem{van2003workflow}
{Wil~MP van~der Aalst}, {Boudewijn~F van Dongen}, {Joachim Herbst}, {Laura
  Maruster}, {Guido Schimm}, {and} {AJMM Weijters}. 2003.
\newblock \showarticletitle{Workflow Mining: a survey of issues and
  approaches}.
\newblock {\em Data \& Knowledge engineering\/} {47}, 2 (2003), 237--267.
\newblock


\bibitem{viegas_studying_2004}
{Fernanda~B. Viégas}, {Martin Wattenberg}, {and} {Kushal Dave}. 2004.
\newblock \showarticletitle{Studying {Cooperation} and {Conflict} {Between}
  {Authors} with {History} {Flow} {Visualizations}}. In {\em Proc. CHI 2004}.
  ACM, 575--582.
\newblock
\showDOI{%
\url{http://dx.doi.org/10.1145/985692.985765}}


\bibitem{xing_brief_2010}
{Zhengzheng Xing}, {Jian Pei}, {and} {Eamonn Keogh}. 2010.
\newblock \showarticletitle{A {Brief} {Survey} on {Sequence} {Classification}}.
\newblock {\em SIGKDD Explorations Newsletter\/} {12}, 1 (2010), 40--48.
\newblock


\bibitem{yamauchi_collaboration_2000}
{Yutaka Yamauchi}, {Makoto Yokozawa}, {Takeshi Shinohara}, {and} {Toru Ishida}.
  2000.
\newblock \showarticletitle{Collaboration with {Lean} {Media}: {How}
  {Open}-source {Software} {Succeeds}}. In {\em Proc. {CSCW} 2000}. ACM,
  329--338.
\newblock
\showDOI{%
\url{http://dx.doi.org/10.1145/358916.359004}}


\bibitem{yasseri_dynamics_2012}
{Taha Yasseri}, {Robert Sumi}, {Andr\'{a}s Rung}, {Andr\'{a}s Kornai}, {and}
  {János Kert\'{e}sz}. 2012.
\newblock \showarticletitle{Dynamics of {Conflicts} in {Wikipedia}}.
\newblock {\em PLoS ONE\/} {7}, 6 (2012), e38869.
\newblock
\showDOI{%
\url{http://dx.doi.org/10.1371/journal.pone.0038869}}


\bibitem{yoo_digital_2012}
{Yongjin Yoo}. 2012.
\newblock \showarticletitle{Digital Materiality and the Emergence of an
  Evolutionary Science of the Artificial}.
\newblock In {\em Materiality and Organizing: Social Interaction in a
  Technological World}, {Paul~M. Leonardi}, {Bonnie~A. Nardi}, {and} {Jannis
  Kallinikos} (Eds.). Oxford University Press.
\newblock


\end{thebibliography}
\end{document}